\PassOptionsToPackage{breaklinks=true}{hyperref}
\documentclass{lmcs}
\pdfoutput=1

\usepackage{lastpage}
\lmcsdoi{18}{4}{3}
\lmcsheading{}{\pageref{LastPage}}{}{}%
{Dec.~31,~2021}{Nov.~09,~2022}{}

\usepackage[utf8]{inputenc}

\usepackage{amsmath,amssymb,mathtools}
\usepackage{microtype}
\usepackage{colonequals}
\usepackage{bm}
\usepackage{xcolor}
\usepackage{amsmath,amssymb,mathtools}
\usepackage{tikz-qtree}
\usepackage{booktabs}
\usepackage{tabularx}
\usetikzlibrary{automata}
\usepackage{chngcntr}
\usepackage{xspace}
\usepackage{tikz}
\usetikzlibrary{positioning, fit, calc, shapes, arrows, arrows.meta}
\usepackage{algorithm}
\usepackage[noend]{algpseudocode}
\usepackage{cite}

\allowdisplaybreaks


\usepackage{amssymb}
\usepackage{cite}
\usepackage{calc}
\usepackage{tikz}
\usepackage{multicol}
\usetikzlibrary{positioning,chains,fit,shapes,calc}
\usepackage{subcaption}

\makeatletter
\newcommand*{\defeq}{\mathrel{\rlap{%
  \raisebox{0.3ex}{$\m@th\cdot$}}%
  \raisebox{-0.3ex}{$\m@th\cdot$}}%
  =}
\makeatother

\newcommand{\card}[1]{\left|{#1}\right|}

\renewcommand{\phi}{\varphi}

\renewcommand{\Pr}{\mathrm{Pr}}
\renewcommand{\S}{\mathrm{S}}

\newcommand{\node}{\mathsf{node}}

\newcommand{\UMC}{\mathsf{UR}}
\newcommand{\WUR}{\mathsf{WUR}}
\newcommand{\PQE}{\mathsf{PQE}}

\newcommand{\ccc}{\mathrm{c}}
\newcommand{\dd}{\mathrm{d}}
\newcommand{\ee}{\mathrm{e}}

\def\e#1{\emph{#1}}
\def\scs{\mathbf{S}}

\def\const{\mathsf{Const}}

\def\tup#1{\vec{#1}}
\def\dl{\mathrel{{:}{\text{-}}}}

\usepackage{mathtools} \def\set#1{\mathord{\{#1\}}}
\def\modelset{\mathrm{Mod}} \def\prob{\mathrm{Pr}}

\def\atoms{\mathrm{atoms}}

\def\RST{Q_{1}}

\def\R{\mathsf{R}}
\def\S{\mathsf{S}}
\def\T{\mathsf{T}}

\def\D{\mathcal{D}}

\newcommand{\notR}{\not\!\R}
\newcommand{\notS}{\not\!\S}
\newcommand{\notT}{\not\!\T}

\newtheorem{commentthm}[thm]{Comment}
\newenvironment{comment}
{\begin{commentthm}\em}
{\qed\end{commentthm}}

\begin{document}

\title[Uniform Reliability of Self-Join-Free Conjunctive Queries]{Uniform
Reliability of Self-Join-Free\texorpdfstring{\\}{} Conjunctive Queries}

\author[A.~Amarilli]{Antoine Amarilli\lmcsorcid{0000-0002-7977-4441}\rsuper{a}}
\author[B.~Kimelfeld]{Benny Kimelfeld\lmcsorcid{0000-0002-7156-1572}\rsuper{b}}

\address{LTCI, T\'el\'ecom Paris, Institut Polytechnique de Paris, France}
\email{antoine.amarilli@telecom-paris.fr}

\address{Technion - Israel Institute of Technology, Haifa, Israel}
\email{bennyk@cs.technion.ac.il}

\begin{abstract}
 The \emph{reliability} of a Boolean Conjunctive Query (CQ) over a
 tuple-independent probabilistic database is the probability that the
 CQ is satisfied when the tuples of the database are sampled one by
 one, independently, with their associated probability.  For queries
 without self-joins (repeated relation symbols), the data complexity
 of this problem is fully characterized by a known dichotomy:
 reliability can be computed in polynomial time for
 \emph{hierarchical} queries, and is \#P-hard for non-hierarchical
 queries.

 Inspired by this dichotomy, we investigate a fundamental counting
 problem for CQs without self-joins: how many sets of facts from the
 input database satisfy the query? This is equivalent to the
 \emph{uniform} case of the query reliability problem, where the
 probability of every tuple is required to be~$\frac 12$.  Of course, for
 hierarchical queries, uniform reliability is solvable in polynomial
 time, like the reliability problem. We show that being hierarchical
 is also necessary for this tractability (under conventional
 complexity assumptions). In fact, we establish a generalization of
 the dichotomy that covers every restricted case of reliability in which the
  probabilities of tuples are determined by their relation.
\end{abstract}

\maketitle

\section{Introduction}
\label{sec:intro}

\e{Probabilistic databases}~\cite{DBLP:series/synthesis/2011Suciu} extend the usual model of relational databases by allowing database facts to be uncertain, in order to model noisy and imprecise data.  The evaluation of a Boolean query $Q$ over a probabilistic database~$D$ is then the task of computing the probability that $Q$ is true under the probability distribution over possible worlds given by~$D$.  This computational task has been considered by Gr{\"{a}}del, Gurevich and Hirsch~\cite{DBLP:conf/pods/GradelGH98} as a special case of computing the \e{reliability} of a query in a model which is nowadays known as \e{Tuple-Independent probabilistic Databases} (TIDs)~\cite{DBLP:series/synthesis/2011Suciu,dalvi2007efficient}.  In a TID, every fact is associated with a probability of being true, and the truth of every fact is an independent random event.  While the TID model is rather weak, query evaluation over TIDs can also be used for probabilistic inference over models with correlations among facts, such as Markov Logic Networks~\cite{DBLP:journals/pvldb/GribkoffS16,DBLP:journals/pvldb/JhaS12}.  Hence, studying the complexity of query evaluation on TIDs is the first step towards understanding which forms of probabilistic data can be tractably queried.

To this end, Gr{\"{a}}del et al.~\cite{DBLP:conf/pods/GradelGH98} showed the first Boolean Conjunctive Query (referred to simply as a \e{CQ} hereafter) for which query evaluation is \#P-hard on TIDs.  Later, Dalvi and Suciu~\cite{dalvi2007efficient} established a dichotomy on the complexity of evaluating CQs without self-joins (i.e., without repeated relation symbols) over TIDs: if the CQ is \e{safe} (or \e{hierarchical}~\cite{DBLP:journals/cacm/DalviRS09,DBLP:series/synthesis/2011Suciu} as we explain next), the problem is solvable in polynomial time; otherwise, the problem is \#P-hard.  (This result was later extended to the class of all CQs and unions of CQs~\cite{dalvi2012dichotomy}.)

The class of \e{hierarchical} CQs is defined by requiring that, for every two variables $x$ and $y$, the sets of query atoms that feature $x$ must contain, be contained in, or be disjoint from, the set of atoms that feature $y$.  Beyond query evaluation on TIDs, this class of hierarchical queries was found to characterize the tractability boundary of other query evaluation tasks for CQs without self-joins, over databases without probabilities (and under conventional complexity assumptions). Olteanu and Huang~\cite{DBLP:conf/sum/OlteanuH08} showed that a query is hierarchical if and only if, for every database, the \e{lineage} of the query is a read-once formula. Livshits, Bertossi, Kimelfeld and Sebag~\cite{DBLP:conf/icdt/LivshitsBKS20} proved that the hierarchical CQs are precisely the ones that have a tractable \e{Shapley value} as a measure of responsibility of facts to query answers (a result that was later generalized to CQs with negation~\cite{DBLP:conf/icdt/LivshitsBKS20}); they also conjecture that this complexity classification also holds for another measure of responsibility, namely the \e{causal effect}~\cite{DBLP:conf/tapp/SalimiBSB16}.  (We discuss these measures again later in this introduction.)  Berkholz, Keppeler and Schweikardt~\cite{DBLP:conf/pods/BerkholzKS17} showed that the hierarchical CQs\footnote{For clarification, the tractability condition of
  Berkholz et al.~\cite{DBLP:conf/pods/BerkholzKS17} is called ``q-hierarchical'' and it is a strict restriction of the condition of being hierarchical for non-Boolean conjunctive queries. As they explain, the two properties coincide in the Boolean case (i.e., the case discussed here), that is, a Boolean CQ is q-hierarchical if and only it is hierarchical.}
  are (up to conventional assumptions of fine-grained complexity) precisely the ones (Boolean) CQs for which we can use an auxiliary data structure to update the query answer in constant time in response to the insertion or deletion of a tuple.

In this paper, we show that the property of being hierarchical also captures the complexity of a fundamental counting problem for CQs without self-joins: \e{how many sets of facts from the input database satisfy the query?}  This problem, which we refer to as \e{uniform reliability}, is equivalent to query evaluation over a TID where the probability of \e{every} fact is equal to~$\frac12$.  In particular, it follows from the aforementioned dichotomy that this problem can be solved in polynomial time for every self-join-free hierarchical CQ $Q$.  Yet, if $Q$ is not hierarchical, it does not necessarily mean that $Q$ is intractable already in this uniform setting. Indeed, it was not known whether enforcing uniformity makes query evaluation on TIDs easier, and the complexity of uniform reliability was already open for the simplest case of a non-hierarchical CQ: $\RST\dl \R(x),\S(x,y),\T(y)$.  The proofs of \#P-hardness of Dalvi and Suciu~\cite{dalvi2007efficient} require TIDs with deterministic facts (probability $1$), in addition to $\frac12$, already in the case of~$\RST$.  Here, we address this problem and show that the dichotomy is also true for the uniform reliability problem.  In particular, uniform reliability is \#P-complete for every non-hierarchical CQ without self-joins (and solvable in polynomial time for every hierarchical CQ without self-joins). In fact, we establish a more general result for the problem of \e{weighted uniform reliability}, that we discuss later on.

The uniform reliability problem that we study is a basic combinatorial problem on CQs, and a natural restricted case of query answering on TIDs, but it also has a direct application for quantifying the impact (or responsibility) of a fact $f$ on the result of a CQ~$Q$ over ordinary (non-probabilistic) databases. One notion of tuple impact is the aforementioned causal effect, defined as the difference between two quantities: the probability of $Q$ conditioning on the existence of~$f$, minus the probability of $Q$ conditioning on the absence of $f$~\cite{DBLP:conf/tapp/SalimiBSB16}.  This causal effect was recently shown~\cite{DBLP:conf/icdt/LivshitsBKS20} to be the same as the \e{Banzhaf power index}, studied in the context of wealth distribution in cooperative game theory~\cite{10.2307/3689345} and applied, for instance, to voting in the New York State Courts~\cite{Grofman1979}.  One notion of causal effect (with so-called \e{endogenous} facts) is defined by viewing the ordinary database as a TID where the probability of every fact is $\frac12$.  Therefore, computing the causal effect amounts to solving two variations of uniform reliability, corresponding to the two quantities.  In fact, it is easy to see that all of our results apply to each of these two variations.

Uniform reliability also relates to the aforementioned computation of a tuple's Shapley value, a measure of wealth distribution in cooperative game theory that has been applied to many use cases~\cite{shapley53,roth1988shapley}.  Livshits et al.~\cite{DBLP:conf/icdt/LivshitsBKS20} showed that computing a tuple's Shapley value can be reduced to a generalized variant of uniform reliability. Specifically, for CQs, computing the Shapley value (again for \e{endogenous} facts) amounts to calculating the number of subinstances that satisfy $Q$ \e{and} have precisely $m$ tuples (for a given number $m$). This generalization of uniform reliability is tractable for every hierarchical CQ without self-joins~\cite{DBLP:conf/icdt/LivshitsBKS20}. Clearly, our results here imply that this generalization is intractable for every non-hierarchical CQ without self-joins, allowing us to conclude that the dichotomy  in complexity also applies to this generalization.

Some natural generalizations of uniform reliability lie between model counting and probabilistic query answering.  These include the case where the probability of each tuple of the database is the same, but not necessarily $\frac12$.  This problem can arise, for example, in scenarios of network reliability, where all connections are equally important and have the same independent probability of failure.  A more general case is the one where the probabilities for every relation are the same, but different relations may be associated with different probabilities. This corresponds to data integration scenarios where every relation is a resource with a different level of trust (e.g., enterprise data vs.~Web data vs.~noisy sensor data). The latter variation is the one we refer to as \emph{weighted uniform reliability}.

Our result in its full generality (namely Theorem~\ref{thm:hardgeneral_Q}, and its generalization to deterministic queries, Theorem~\ref{thm:deterministic}) completely determines the complexity of weighted uniform reliability: for every non-hierarchical CQ without self-joins, and for \e{every} fixed assignment of probabilities to relations, probabilistic query answering is \#P-hard when the fixed probabilities are less than~$1$ (while for every hierarchical CQ without self-joins the problem is solvable in polynomial time, as already known due to Dalvi and Suciu~\cite{dalvi2007efficient}). When the fixed probabilities can be~$1$, we have a more complex classification inspired by \cite[Theorem 8]{dalvi2007efficient}, shown as Theorem~\ref{thm:deterministic}.

\paragraph*{Related work.}  As explained earlier, our work is closely related to existing literature on query evaluation over probabilistic databases.  The dichotomy of Dalvi and Suciu~\cite{dalvi2007efficient} for CQs without self-joins requires tuples with probabilities $\frac12$ and $1$. This is also the case for their generalized dichotomy on CQs without self-joins, which covers deterministic relations (i.e., all tuples have probability $1$) but allows tuples in the remaining relations to have arbitrary probabilities (including~1).  The later generalization of the dichotomy by Dalvi and Suciu~\cite{dalvi2012dichotomy} to CQs with self-joins and to UCQs required an unbounded class of probabilities, not just~$\frac12$ and~$1$.  In very recent work, Kenig and Suciu~\cite{DBLP:journals/corr/abs-2008-00896} have strengthened the generalized dichotomy and showed that probabilities $\frac12$ and $1$ suffice for UCQs as well.\footnote{Kenig and Suciu refer to this case as TID with probabilities from $\set{0,\frac 12,1}$; we mean the same thing, as in this paper we assume that tuples with probability zero are simply ignored.}  In that work, they also investigate uniform reliability (that we study here, i.e., where $\frac12$ is the only nonzero probability allowed) and prove \#P-hardness for the so-called unsafe ``final type-I'' queries. As they explain in their discussion on the work of this paper (which was posted as a preprint before theirs), their result on uniform reliability complements ours, and it is not clear if any of these two results can be used to prove the other.

Our work in this paper also relates to rewriting techniques used in the case of DNF formulas to reduce weighted model counting to unweighted model counting~\cite{chakraborty2015weighted}. Nevertheless, the results and techniques for this problem are not directly applicable to ours, since model counting for CQs translates to DNFs of a very specific shape (namely, those that can be obtained as the lineage of the query).

Another superficially related problem is that of \e{symmetric model counting}~\cite{beame2015symmetric}. This is a variant of uniform reliability where each relation consists of \emph{all possible tuples} over the corresponding domain, and so each fact carries the same weight: these assumptions are often helpful to make model counting tractable. The assumption that we make on databases is much weaker: we do not deal with symmetric databases, but rather with arbitrary databases where all facts of the database (but \emph{not} necessarily all possible facts over the domain) have the same probability, or have a common probability defined by the relation, when they are present.  For this reason, the tractability results of Beame et al.~\cite{beame2015symmetric} do not carry over to our setting. In terms of hardness results, \cite[Theorem~3.1]{beame2015symmetric} shows the \#P$_1$-hardness of symmetric model counting (hence of uniform reliability) for a specific FO$^3$ sentence, and \cite[Corollary~3.2]{beame2015symmetric} shows a \#P$_1$-hardness result for \emph{weighted} symmetric model counting for a specific CQ (without assuming self-join-freeness). Hence, these results do not determine the complexity of uniform reliability for self-join-free CQs as we do here.

There is a closer connection to existing dichotomy results on counting \e{database repairs}~\cite{DBLP:journals/jcss/MaslowskiW13,DBLP:conf/icdt/MaslowskiW14}.  In this setting, the input database may violate the primary key constraints of the relations, and a repair is obtained by selecting one fact from every collection of conflicting facts (i.e., distinct facts that agree on the key): the \e{repair counting problem} asks how many such repairs satisfy a given CQ. In particular, it can easily be shown that for a CQ $Q$, there is a reduction from the uniform reliability of $Q$ to repair counting of another CQ $Q'$. Yet, this reduction can only explain cases of tractability (namely, where $Q$ is hierarchical) which, as explained earlier, are already known.  We do not see how to design a reduction in the other direction, from repair counting to uniform reliability, in order to show our hardness result.

Finally, our work relates to the study of the Constraint Satisfaction Problem (CSP). However, there are two key differences.  First, we study query evaluation in terms of homomorphisms \emph{from} a fixed CQ, whereas the standard CSP phrasing talks about homomorphisms \emph{to} a given template. Second, the standard counting variant of CSP (namely, \#CSP), for which Bulatov has proved a dichotomy~\cite{DBLP:journals/jacm/Bulatov13}, is about counting \emph{the number of homomorphisms}, whereas we count \emph{the number of subinstances} for which a homomorphism exists. For these reasons, it is not clear how results on CSP and \#CSP can be helpful towards our main result.

\paragraph*{Prior publication.} A short version of this manuscript appeared in conference proceedings~\cite{DBLP:conf/icdt/AmarilliK21}.
In terms of the results, the main difference between the versions is that this manuscript establishes a stronger result, that is, a dichotomy for \e{weighted} uniform reliability (with arbitrary probabilities per relation) rather than unweighted uniform reliability (with the single probability $\frac12$). In fact, the question of weighted uniform reliability has been stated as an open problem in the conference publication, and we have posed a conjecture regarding the query $\RST$~\cite[Conjecture~7.4] {DBLP:conf/icdt/AmarilliK21}. We prove this conjecture in the present work. We do so by showing our result on weighted uniform reliability for arbitrary queries when we do not allow deterministic relations (Sections~\ref{sec:reduction1}--\ref{sec:invertible}), and addressing the case of deterministic relations in Section~\ref{sec:deterministic}.
Interestingly, our stronger result allows for a simpler proof structure, since now it is possible to define a reduction from $\RST$ to every non-hierarchical CQ (while the probabilities may be different).
Besides the stronger results, compared to the conference version, this manuscript includes complete proofs.

\paragraph*{Organization.}  We give preliminaries in Section~\ref{sec:prelim}. In Section~\ref{sec:result}, we formally state the studied problems and main results, that is, the dichotomy on the complexity of uniform reliability, and more generally weighted uniform reliability, for CQs without self-joins. We prove this result in Sections~\ref{sec:reduction1}--\ref{sec:invertible}.  We discuss a generalization that allows for deterministic relations in Section~\ref{sec:deterministic}, and conclude in Section~\ref{sec:conc}.

\section{Preliminaries}
\label{sec:prelim}

We begin with some preliminary definitions and notation that we use
throughout the paper. We first define databases and conjunctive queries, before
introducing the task of probabilistic query evaluation, and the uniform
reliability problem that we study.

\paragraph*{Databases.} A (relational) \e{schema} $\scs$ is a
collection of \e{relation symbols} with each relation symbol $U$ in
$\scs$ having an associated arity.  We assume a countably infinite set
$\const$ of \e{constants} that are used as database values.  A
\e{fact} over $\scs$ is an expression of the form
$U(c_1,\dots,c_k)$ where $U$ is a relation symbol of $\scs$,
where $k$ is the arity of~$U$, and where $c_1,\dots,c_k$ are values
of $\const$. An \e{instance}~$I$ over $\scs$ is a finite set of facts.
In particular, we say that an instance $J$ is a \e{subinstance} of an instance
$I$ if we have $J \subseteq I$.

\paragraph*{Conjunctive queries.}  This paper focuses on queries in
the form of a Boolean Conjunctive Query, which we refer to simply as a
\e{CQ}.  Intuitively, a CQ $Q$ over the schema $\scs$ is a relational
query definable as an existentially quantified conjunction of atoms.
  Formally, a CQ is a first-order formula of the form
$Q \dl U_1(\tup{a}_1),\dots,U_n(\tup{a}_m)$ where each
$U_i(\tup{a}_i)$ is an \e{atom} of $Q$, formed of a relation
symbol of $\scs$ and of a tuple $\tup{a}_i$ of constants and (existentially
quantified) variables, with the same arity as
$U_i$. In the context of a CQ $Q$, we omit the schema $\scs$ and
implicitly assume that $\scs$ consists of the relation symbols that
occur in $Q$ (with the arities that they have in $Q$); in that case,
we may also refer to an instance $I$ over $\scs$ as an instance
\e{over $Q$}.  We write $I\models Q$ to state that the instance $I$
satisfies $Q$.  We denote the set of all subinstances $J$ of $I$
that satisfy $Q$ by:
\[\modelset(Q,I) \defeq \set{J\subseteq I\mid J\models Q}.\]

A \e{self-join} in a CQ $Q$ is a pair of distinct atoms over the same
relation symbol. For example, in $Q \dl \R(x,y),\S(x),\R(y,z)$, the
first and third atoms constitute a self-join. Our analysis in this
paper is restricted to CQs \e{without self-joins}, that we also call
\e{self-join-free}.

Let $Q$ be a CQ. For each variable $x$ of $Q$, we denote by
$\atoms(x)$ the set of atoms $U_i(\tup{\tau}_i)$ of $Q$ where $x$
occurs.  We say that $Q$ is \e{hierarchical}~\cite{dalvi2007efficient}
if for all variables $x$ and $x'$ one of the following three relations
hold: $\atoms(x)\subseteq \atoms(x')$, $\atoms(x')\subseteq
\atoms(x)$, or $\atoms(x)\cap \atoms(x')=\emptyset$.  The simplest
non-hierarchical self-join-free CQ is $\RST$, which we already mentioned in the
introduction:
\begin{equation}\label{eq:QRST}
\RST \dl \R(x),\S(x,y),\T(y)
\end{equation}

\paragraph*{Probabilistic query evaluation.}  The problem of
\e{probabilistic query evaluation} over tuple-independent
databases~\cite{DBLP:series/synthesis/2011Suciu} is defined as
follows.

\begin{defi}
  The problem of \emph{probabilistic query evaluation} (or PQE) for a
  CQ $Q$, denoted $\PQE(Q)$, is that of computing, given an instance
  $I$ over $Q$ and an assignment $\pi:I\rightarrow[0,1]$ of a
  probability $\pi(f)$ to every fact $f$, the probability that $Q$ is
  true, namely:
\[\prob(Q,I,\pi)\defeq \sum_{J\in\modelset(Q,I)}\prod_{f\in J}\pi(f)\times
\prod_{f\in I\setminus J}(1-\pi(f))\,.\]
\end{defi}
We again study the \emph{data complexity} of this problem, and we assume that
the probabilities attached to the instance~$I$ are rational numbers represented by their
integer numerator and denominator.

PQE was first studied by Gr{\"{a}}del,
Gurevich and Hirsch~\cite{DBLP:conf/pods/GradelGH98} as
\e{query reliability} (which they also generalize beyond Boolean queries).
They identified a Boolean CQ $Q$ with self-joins such that the reliability of $Q$ is
\#P-hard to compute.
Dalvi and Suciu~\cite{dalvi2007efficient,dalvi2012dichotomy} then
studied the PQE problem, culminating in their dichotomy for the
complexity of PQE on unions of conjunctive queries with
self-joins~\cite{dalvi2012dichotomy}.  In this paper, we only
consider their earlier study of CQs without
self-joins~\cite{dalvi2007efficient}.
They characterize, under conventional complexity assumptions, the self-join-free
CQs where PQE is solvable in PTIME. They state the result in terms of safe query
plans (``safe CQs''), but the term ``hierarchical'' was adopted in later
  publications~\cite{DBLP:journals/cacm/DalviRS09,DBLP:series/synthesis/2011Suciu}:

\begin{thmC}[\cite{dalvi2007efficient}]\label{thm:DS-dichotomy}
Let $Q$ be a CQ without self-joins. If $Q$ is hierarchical,
then $\PQE(Q)$ is solvable in polynomial time. Otherwise,
$\PQE(Q)$ is \#P-hard.
\end{thmC}

Recall that \#P is the complexity class of problems that count witnesses of
an NP-relation (e.g., satisfying assignments of a logical formula,
vertex covers of a graph, etc.).  A function $F$ is \#P-hard if every
function in \#P has a polynomial-time \e{Turing reduction} (or \e{Cook
  reduction}) to~$F$.

We stress that Theorem~\ref{thm:DS-dichotomy} applies to CQs
\e{without} self-joins.  In the presence of self-joins,
being hierarchical is still necessary for tractability, but no
longer sufficient~\cite[Theorem~4.23, Proposition~4.25]{DBLP:series/synthesis/2011Suciu}.

\section{Problem Statement and Main Result}\label{sec:result}
We study the query reliability problem (which we equivalently refer to
as PQE). Our main focus is on the uniform variant of this problem, where the
probability of every fact is $\frac12$. Equivalently, the task is to
count the subinstances that satisfy the query (up to
division/multiplication by $2^n$ where $n$ is the number of facts in
the instance). Formally:

\begin{defi}
  The problem of \emph{uniform reliability} for a CQ $Q$, denoted
  $\UMC(Q)$, is that of determining, given an instance $I$ over $Q$,
  how many subinstances of $I$ satisfy $Q$. In other words, $\UMC(Q)$
  is the problem of computing $\card{\modelset(Q,I)}$ given $I$. We study the
  \e{data complexity} of this problem, i.e., $Q$ is fixed and the complexity is
  a function of the input~$I$.
\end{defi}

Let $Q$ be a CQ without self-joins.  It follows from
Theorem~\ref{thm:DS-dichotomy} that, if $Q$ is hierarchical, then
$\UMC(Q)$ is solvable in polynomial time. Indeed, there is a
straightforward reduction from $\UMC(Q)$ to $\PQE(Q)$: given an
instance $I$ for $Q$, let $\pi:I\rightarrow[0,1]$ be the function that
assigns to every fact $f$ of $I$ the probability $\pi(f)=\frac12$. Then we
have:
\[\card{\modelset(Q,I)}=2^{|I|}\times \prob(Q,I,\pi)\] because every
subset of $I$ has the same probability, namely $2^{-|I|}$.

However, the other direction is not evident. If $Q$ is
non-hierarchical, we know that $\PQE(Q)$ is \#P-hard, but we do not
know whether the same is true of $\UMC(Q)$. Indeed, this does not
follow from Theorem~\ref{thm:DS-dichotomy} (as uniform reliability is
a restriction of PQE), and it does not follow from the proof of the
theorem either. Specifically, the reduction that Dalvi and
Suciu~\cite{dalvi2007efficient} used to show hardness consists of two
steps.
\begin{enumerate}
\item Proving that $\PQE(\RST)$ is \#P-hard (where $\RST$ is
  defined in~\eqref{eq:QRST}).
\item Constructing a polynomial-time many-one reduction from $\PQE(\RST)$ to $\PQE(Q)$
  for every non-hierarchical CQ $Q$ without self-joins. 
\end{enumerate}
In both steps, the constructed instances $I$ consist of facts with two probabilities: $\frac12$ and $1$ (i.e., \emph{deterministic facts}). If all facts had probability $\frac12$, then we would get a reduction to our $\UMC(Q)$ problem. However, the proof crucially relies on deterministic facts, and we do not see how to modify it to give the probability $\frac12$ to all facts. This is true for both steps. Even for the first step, the complexity of $\UMC(\RST)$ has been unknown so far. For the second step, it is not at all clear how to reduce from $\UMC(\RST)$ to $\UMC(Q)$, even if $\UMC(\RST)$ is proved to be \#P-hard.

In this paper, we resolve the question and prove that $\UMC(Q)$ is \#P-complete whenever $Q$ is a non-hierarchical CQ without self-joins.  Hence, we establish that the dichotomy of Theorem~\ref{thm:DS-dichotomy} also holds for uniform reliability.  Our main result is:

\begin{thm}\label{thm:UMC-dichotomy}
  Let $Q$ be a CQ without self-joins. If $Q$ is hierarchical, then $\UMC(Q)$ is solvable in polynomial time. Otherwise, $\UMC(Q)$ is \#P-complete.
\end{thm}

As said above, the hardness side of Theorem~\ref{thm:UMC-dichotomy} is
essentially the statement that when $Q$ is non-hierarchical, $\PQE(Q)$
is \#P-hard even when every fact has the probability $\frac12$, and
this is what we need to prove. In fact, we prove
something stronger: Fix any CQ $Q$ without self-joins, assign to
each relation symbol $U$ of $Q$ an arbitrary probability
$\phi(U)$ (which may be different from $\frac12$), and consider
$\PQE(Q)$ on instances $I$ where fact probabilities are given by~$\phi$.
We refer to this problem
as the \e{weighted uniform reliability} problem, and write it $\WUR(Q, \phi)$.
Formally:

\begin{defi}
  \label{def:pqebis}
  Let $Q$ be a CQ without self-joins, and let $\phi$ be a function
  mapping each relation symbol $U$ of~$Q$ to a rational number
  $0 < \phi(U) \leq 1$.  The \emph{weighted uniform reliability}
  problem $\WUR(Q, \phi)$ is the problem of probabilistic query
  evaluation $\PQE(Q)$ for~$Q$ on input instances whose probability
  function $\pi: I \to [0, 1]$ is defined by~$\phi$, that is, for
  every fact $f \in I$, we have $\pi(f) = \phi(U)$ where $U$ is the
  relation symbol of $f$.
\end{defi}

The problem $\WUR(Q, \phi)$ is tractable if $Q$ is hierarchical (because it is a
special case of PQE). If $Q$ is
non-hierarchical, then $\PQE(Q)$ is \#P-hard. In this paper,
we show that, when all relation probabilities are strictly less than~$1$, the
problem $\WUR(Q, \phi)$ is also intractable.

\begin{comment}
  Note that it is not clear, to begin with, that $\PQE(Q)$
    remains hard if we restrict the use of deterministic facts (and,
    in fact, increase the level of uncertainty).  To illustrate that,
    we recall an example from the ``Probabilistic Databases''
    book~\cite{DBLP:series/synthesis/2011Suciu}: 
    consider $\PQE(Q_1)$ for $Q_1$ the hard query of Equation~\ref{eq:QRST},
    under the restriction that $S$ is the cartesian product of the relations $R$
    and $T$, and that every tuple has probability $\frac 12$. In other words,  this
    is the problem of computing the probability 
    that a \emph{complete} bipartite graph contains an edge and its two incident
    vertices, when every vertex and edge can independently disappear
    with probability $\frac 12$.
    This problem is solvable in polynomial time by a fairly simple counting 
    argument~\cite[page 47]{DBLP:series/synthesis/2011Suciu}.
    Nevertheless, when one reduces the uncertainty by allowing some tuples to
    have probabilities $0$ or~$1$, then the problem becomes hard, as $\PQE(Q_1)$
    is \#P-hard when the tuples can have probabilities $0$, $\frac 12$, or $1$.
    Hence, in this situation, a problem can become tractable when the use of
    deterministic facts is restricted.
\end{comment}

\begin{thm}
  \label{thm:hardgeneral_Q}
  Let $Q$ and $\varphi$ be as in Definition~\ref{def:pqebis}. If $Q$
  is non-hierarchical, and if $\varphi$ maps each relation symbol to a
  probability strictly less than~$1$, then $\WUR(Q, \phi)$ is \#P-hard.
\end{thm}

Note that, even though uniform reliability is a special case of weighted uniform
reliability, Theorem~\ref{thm:hardgeneral_Q} is finer than
Theorem~\ref{thm:UMC-dichotomy}, and does not follow from it. Indeed,
Theorem~\ref{thm:hardgeneral_Q} implies that each (fixed) choice of probability
is intractable, in particular the choice giving probability $\frac12$ to every
relation.

We present the proof of this result in most of this paper
(Sections~\ref{sec:reduction1}--\ref{sec:invertible}), before discussing in
Section~\ref{sec:deterministic} a generalization where probability~$1$ is allowed.

\section{Reducing from $Q_1$ to Arbitrary Queries}
\label{sec:reduction1}

The structure of our proof of Theorem~\ref{thm:hardgeneral_Q}  is
analogous to (but very different from) the reduction of Dalvi and
Suciu~\cite{dalvi2007efficient}. We prove the \#P-hardness of
$\WUR(Q, \phi)$ in two steps.
\begin{enumerate}
\item Prove that $\WUR(Q_1, \phi)$ is \#P-hard for every $\phi$ (where
  $Q_1$ is defined in~\eqref{eq:QRST});
   \item Prove that for every $Q$ and $\phi$ as in
     Theorem~\ref{thm:hardgeneral_Q} there exists $\phi_1$ such that
     there is a polynomial-time many-one reduction from $\WUR(Q_1,
     \phi_1)$ to $\WUR(Q, \phi)$.
\end{enumerate}

In this section, we present the second step. The first step is much more
challenging, and presented in
Sections~\ref{sec:reduction}--\ref{sec:invertible}. Here is the statement of the
first step:

\begin{lem}
  \label{lemma:reduction_to_rst}
  Let $Q$ and $\phi$ be as in Definition~\ref{def:pqebis}, where $\phi$ maps to
  probabilities strictly less than~$1$. If $Q$ is
  non-hierarchical then there exists $\phi_1$, also mapping to probabilities
  strictly less than~$1$, such that there is a
  polynomial-time many-one reduction from $\WUR(Q_1, \phi_1)$ to
  $\WUR(Q, \phi)$.
  \end{lem}

We prove Lemma~\ref{lemma:reduction_to_rst} using the following, more
elaborate lemma.
\begin{lem}\label{lemma:reduction_detailed}
  Let $Q \dl U_1(\tup{a}_1),\dots,U_n(\tup{a}_m)$ be a self-join-free CQ, and
  let $\phi$ be as in
  Definition~\ref{def:pqebis} and mapping each relation to a probability
  strictly less than~$1$. Suppose that $Q$ is non-hierarchical, and
  let $x$ and $y$ be two variables of $Q$ witnessing this, i.e., $\atoms(x)$ and
  $\atoms(y)$ have a nonempty intersection and none contains the
  other.  Let $\phi_1$ be the mapping defined as follows.
  \begin{itemize}
  \item $\phi_1(\R)$ is the product of the $\phi(U_i)$ over all $i$
    such that $U_i(\tup{a}_i)$ contains $x$ and not $y$;
\item $\phi_1(\S)$ is the product of the $\phi(U_i)$ over all $i$
    such that $U_i(\tup{a}_i)$ contains both $x$ and $y$;
\item $\phi_1(\T)$ is the product of the $\phi(U_i)$ over all $i$
    such that $U_i(\tup{a}_i)$ contains $y$ and not $x$.
\end{itemize}
There is a polynomial-time many-one reduction from $\WUR(Q_1, \phi_1)$ to
  $\WUR(Q, \phi)$.
\end{lem}  
\begin{proof}
  Let $I_1$ be an input instance for $\WUR(Q_1, \phi_1)$. If $(a,b)$ is
  a pair of constants such that $I_1$ contains all of $\R(a)$,
  $\S(a,b)$ and $\T(b)$, then we call $(a,b)$ a \e{match}. Without
  loss of generality, we assume that $I_1$ has no dangling
  tuples, that is, every fact is a part of one or more matches, as we
  can remove all other facts without changing the probability
  of~$Q_1$.

  Fix any constant $c$. For an atom $U_i(\tup{a}_i)$ of $Q$, we denote
  by $f_i(a,b)$ the fact over $U_i$ that is obtained from
  $U_i(\tup{a}_i)$ by replacing every occurrence of $x$ with $a$, every
  occurrence of $y$ with $b$, and every occurrence of every other
  variable with $c$. In particular, if $U_i(\tup{a}_i)$ contains
  neither $x$ nor $y$, then $f_i(a,b)$ contains only $c$ and the
  constants of $\tup{a}_i$.

  We construct the input instance $I$ for $\WUR(Q, \phi)$ by taking
    every match $(a,b)$ from $I_1$ and inserting into $I$ the facts
    $f_i(a,b)$ for all $i=1,\dots,n$. Following the construction, let us define 
    the following for a match $(a,b)$:
  \begin{itemize}
    \item $F_x(a)$ is the set of facts $f_i(a,b)$ inserted
      due to atoms $U_i(\tup{a}_i)$ that contain $x$ and not $y$;
      \item $F_{x,y}(a,b)$ is the set of facts $f_i(a,b)$  inserted
        due to atoms $U_i(\tup{a}_i)$ with both $x$ and $y$;
        \item $F_y(b)$ is the set of facts $f_i(a,b)$ inserted
      due to atoms $U_i(\tup{a}_i)$ that contain $y$ and not $x$;
      \item $F()$ is the set of facts $f_i(a,b)$ inserted due
        to atoms $U_i(\tup{a}_i)$ that contain neither $x$ nor $y$.
        Note that, by construction, these facts only contain the constants
        of~$Q$ and the constant~$c$; hence, for each such atom $U_i$, there is
        precisely one fact of this relation in~$I$.
    \end{itemize}

    We complete the proof by showing
    the following:
  \begin{equation}\label{eq:reduction}
    \prob(Q,I,\pi)=\prob(Q_1,I_1,\pi_1)\times \prod_{f\in F()}\pi(f)
  \end{equation}
  where $\pi$ and $\pi_1$ are the probability functions defined respectively
  by~$\phi$ and~$\phi_1$ as in Definition~\ref{def:pqebis}.

  To show this, let us first observe that possible worlds of~$I$ that do not
  contain all facts $F()$ cannot satisfy~$Q$: indeed, each of these facts is the
  only fact for its relation symbol. So this covers the multiplicative factor
  $\prod_{f\in F()}\pi(f)$, and it suffices to consider the possible worlds
  of~$I$ where all facts of~$F()$ are present.

  Let us study these possible worlds of~$I$ by partitioning them
  based on possible worlds of~$I_1$. More precisely, we choose a possible world
  $J$ of~$I$ containing all facts of~$F()$ by first choosing a possible world $J_1 \subseteq I_1$ and:
  \begin{itemize}
    \item for all $a$ in the domain of~$I_1$:
      \begin{itemize}
        \item if $J_1$ contains $\R(a)$, retaining all facts $F_x(a)$
        \item if $J_1$ does not contain $\R(a)$, not retaining at least one of
          the facts of $F_x(a)$
      \end{itemize}
    \item for each match $(a, b)$ in~$I_1$:
      \begin{itemize}
        \item if $J_1$ contains $\S(a,b)$, retaining all facts $F_{x,y}(a,b)$
        \item if $J_1$ does not contain $\S(a,b)$, not retaining at least one of
          the facts of $F_{x,y}(a,b)$
      \end{itemize}
    \item for all $b$ in the domain of~$I_1$:
      \begin{itemize}
        \item if $J_1$ contains $\T(b)$, retaining all facts $F_y(b)$
        \item if $J_1$ does not contain $\T(b)$, not retaining at least one of
          the facts of $F_y(b)$
      \end{itemize}
    \item retaining all facts of~$F()$
  \end{itemize}

  Now, we note that, 
  if a subset $J$ of~$I$ satisfies~$Q$, i.e., there is a
  homomorphism from~$Q$ to~$J$, then the image of the variables~$x$ and~$y$
  determine two elements~$a$ and~$b$ such that $J$ contains all of $F_x(a)$, $F_{x,y}(a,b)$,
  $F_y(b)$ and $F()$, implying in particular that $(a, b)$ is a match of~$Q_1$
  in~$I_1$. Conversely, if~$J$ contains all of these facts for some match $(a,
  b)$ of~$Q_1$ in~$I_1$, then clearly~$J$ satisfies~$Q$. Thus, for every subset
  $J$ of $I$, 
  we have $J\in\modelset(Q,I)$ if and only there is a match
  $(a,b)$ such that $J$ contains all of $F_x(a)$, $F_{x,y}(a,b)$,
  $F_y(b)$ and $F()$. Put differently, when partitioning the possible worlds
  of~$J$ as explained in the previous paragraph, we have 
  $J\in\modelset(Q,I)$ if and
  only if our choice of~$J_1 \subseteq I_1$ satisfies~$Q_1$.
  Hence, we conclude the following.
  \begin{align*}
   \prob(Q,I,\pi) =
    \sum_{J_1\in\modelset(Q_1,I_1)}
    &\left(\prod_{\R(a)\in
    J_1}\prod_{f\in F_x(a)}\pi(f)\right) \left(\prod_{\R(a)\in
      I_1\setminus J_1}\left(1-\prod_{f\in
      F_x(a)}\pi(f)\right)\right)\\
     \times &\left(\prod_{\S(a,b)\in
    J_1}\prod_{f\in F_{x,y}(a,b)}\pi(f)\right) \left(\prod_{\S(a,b)\in
      I_1\setminus J_1}\left(1-\prod_{f\in
      F_{x,y}(a,b)}\pi(f)\right)\right)\\
    \times & \left(\prod_{\T(b)\in
    J_1}\prod_{f\in F_y(b)}\pi(f)\right) \left(\prod_{\T(b)\in
             I_1\setminus J_1}\left(1-\prod_{f\in F_y(b)}\pi(f)\right)\right)\\
              \times &\left(\prod_{f\in F()}\pi(f)\right) 
  \end{align*}
Therefore, from the definition of $\phi_1$, $\phi$, $\pi_1$, and $\pi$
we conclude that:
\begin{align*}
   \prob(Q,I,\pi) =
    \sum_{J_1\in\modelset(Q_1,I_1)}
    &\left(\prod_{\R(a)\in  J_1}\pi_1(\R(a))\right)\left(\prod_{\R(a)\in
    I_1\setminus J_1}(1-\pi_1(\R(a)))\right) \\
     \times &\left(\prod_{\S(a,b)\in
     J_1}\pi_1(\S(a,b))\right)\left(\prod_{\S(a,b)\in I_1\setminus J_1}(1-\pi_1(\S(a,b)))\right) \\
    \times &\left(\prod_{\T(b)\in  J_1}\pi_1(\T(b))\right)\left(\prod_{\T(b)\in
    I_1\setminus J_1}(1-\pi_1(\T(b)))\right) \\
  \times &\left(\prod_{f\in F()}\pi(f)\right)
\end{align*}
Since the first three lines corresponds to simply the probability of
$J_1$, we immediately conclude Equation~\eqref{eq:reduction}, as
promised.  
  \end{proof}

The harder part is the first step, namely that $\WUR(Q_1, \phi)$ is
\#P-hard. We will show it in the next sections. Formally, what remains to
complete the proof of Theorem~\ref{thm:hardgeneral_Q} is to prove the
following:
\begin{thm}
  \label{thm:Qrst-hard}
  Consider the query $Q_1$, and let $\phi$ be a function mapping the
  relation symbols $\R$, $\S$, and $\T$ to rational values such that $0 <
  \phi(\R), \phi(\T) < 1$ and $0 < \phi(\S) \leq 1$.
  Then $\WUR(Q_1, \phi)$ is \#P-hard.
\end{thm}
We will prove Theorem~\ref{thm:Qrst-hard} in
Sections~\ref{sec:reduction}--\ref{sec:invertible}.  Note that this
result does not cover the case where relations have deterministic
facts, except that we do not need to assume that $\phi(\S) < 1$: we
will come back to this issue in Section~\ref{sec:deterministic}.
For now, we only mention that the requirement of being
  strictly smaller than~$1$ is necessary for the correctness of the
  theorem, since the problem is solvable in polynomial time if
  $\phi(\R)=1$ or $\phi(\T)=1$. Lemma~\ref{lemma:reduction_detailed}
  and Theorem~\ref{thm:Qrst-hard} also show where in
  Theorem~\ref{thm:hardgeneral_Q} we are using the assumption 
  that $\varphi$ maps each relation symbol to a probability strictly
  less than~$1$: we need the relevant products of probabilities to be
  smaller than~$1$ in order to be able to reduce from a hard
  configuration of the evaluation of $Q_1$.

\section{Defining the Main Reduction}
\label{sec:reduction}

In this section and the two next ones, we prove
Theorem~\ref{thm:Qrst-hard}, on the CQ $Q_1: \R(x), \S(x, y), \T(y)$. 
For simplicity, let us write $\rho \colonequals \phi(\R)$, $\sigma \colonequals \phi(\S)$, and $\tau \colonequals \phi(\T)$
the respective constant probabilities
of~$\R$, $\S$, and $\T$: we have
$0 < \rho, \tau < 1$ and $0 < \sigma \leq 1$.
We construct a Turing reduction from the \#P-hard
problem of counting the independent sets of a bipartite graph. The
input to this problem is a bipartite graph $G = (R \cup T, S)$ where
$S \subseteq R \times T$, and the goal is to calculate the number $P$
of \emph{independent-set pairs} $(R', T')$ with $R' \subseteq R$ and
$T' \subseteq T$, that is, pairs such that $R' \times T'$ is disjoint
from~$S$.

Counting the independent sets of a bipartite graph is the
  same as computing the number of falsifying assignments of a
  so-called \emph{monotone partitioned 2-DNF formula}, which is a
  monotone Boolean formula in disjunctive normal form over variables
  from two disjoint sets where every clause is the conjunction of one
  variable from one set and one variable from the other. Counting the
  satisfying assignments of such formulas is
  \#P-hard~\cite{provan1983complexity}, so it is also \#P-hard to
  count falsifying assignments, and thus to count independent-set
  pairs. To be more precise, the two sides are viewed as the left and
  right sides, respectively, of the bipartite graph, and every clause
  $x\land y$ corresponds to the edge $(x,y)$. For example, the formula
  \[(x_1\land y_1)\lor (x_1\land y_2)\lor (x_2\land y_2)\] corresponds
  to the bipartite graph $G = (R \cup T, S)$ that has the left side
  $R=\set{x_1,x_2}$, the right side $T=\set{y_1,y_2}$, and the edge
  set $S$ that consists of the pairs $(x_1,y_1)$, $(x_1,y_2)$, and
  $(x_2,y_2)$. A pair $(R',T')$ can be viewed as the assignment of
  truth values that sets to true precisely the variables from $R'$ and
  $T'$. In particular, $(R',T')$ corresponds to a falsifying
  assignment if and only if $R'\cup T'$ is an independent set of the
  graph (i.e., no edge of $S$ has one endpoint in $R'$ and one in
  $T'$). Hence, the number of independent sets is precisely the number
  of falsifying assignments.

\begin{figure}
  \centering
  \input{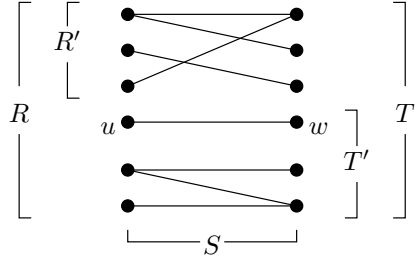}
  \caption{\label{fig:notation} Example of the bipartite graph~$G = (R \cup T, S)$ and an independent set
$(R',T')$}
  \end{figure}

Let us fix $G = (R \cup T, S)$ as the input to the problem.
(See Figure~\ref{fig:notation} for an illustration.)
Our
proof consists of three steps, which we first sketch before presenting
them in detail.

The first step, in the present section, is to describe the
  reduction, that is, how the input bipartite graph~$G$ is used to
  construct in polynomial time instances $D_{p,\phi,\psi}$, for various
  values of~$p$, $\phi$ and $\psi$, on which we invoke our oracle
  for~$\WUR(Q_1, \phi)$ to obtain probabilities $\Pi_{p,\phi,\psi}$.
  The instances $D_{p,\phi,\psi}$ are constructed from~$G$ out of
  building blocks, called \emph{gadgets}: we first introduce them,
  before presenting the construction used in the reduction.

The second step, in Section~\ref{sec:eq-system}, is to show
  that the oracle answers $\Pi_{p,\phi,\psi}$ are related to what the
  reduction needs to compute, that is, the number $P$ of
  independent-set pairs of~$G$. Specifically, we show that we can
  express $P$ as a sum of \emph{variables} of the form
  $X_{i,j,c,d,d'}$. Intuitively, these variables count the subsets of
  the left and right vertices of the bipartite graph satisfying some
  conditions given by the \emph{parameters} $i$, $j$, $c$, $d$ and
  $d'$.  We then show that there is a linear equation system that
  relates these variables to the oracle answers
  $\Pi_{p,\phi,\psi}$. Specifically, we show that there is a vector
  $\vec Y$ defined from these variables which can be expressed by
  multiplying the vector of the $\Pi_{p,\phi,\psi}$ by a square
  matrix~$A'$. We then notice that $A'$ is a Kronecker product of
  three Vandermonde matrices, two of which are easily seen to be
  invertible.

The third and last step of the proof is to show that the matrix $A'$ of
the equation system is invertible, by showing the invertibility of the last
Vandermonde matrix~$A$. This is done in Section~\ref{sec:invertible}, and is the
most technical part of the proof, where we rely on the specific construction of
the instances $D_{p,\phi,\psi}$ and on the gadgets.

\def\lambdaleft{\lambda_{\R}}
\def\barlambdaleft{{\bar\lambda}_{\R}}
\def\lambdaright{\lambda_{\T}}
\def\barlambdaright{{\bar\lambda}_{\T}}
\def\deltaleft{\delta_{\R}}
\def\deltaright{\delta_{\T}}
\def\deltacut{\delta_{\bot}}

\paragraph{Defining the gadgets.}
We start our presentation of the first step by describing the gadgets used in the
reduction as building blocks for our instances 
of~$\WUR(Q_1, \phi)$.
In our reduction, we will use multiple copies of the gadgets, instantiated with
specific elements that will intuitively serve as endpoints to the gadgets.
There are two types of gadgets:

\begin{itemize}
\item The \underline{\emph{$(a,b)$-gadget}} is an instance with two elements $a$
  and $b$ (which are intuitively the endpoints), and the following facts
    (noting that they satisfy the query):
    \[\R(a), \S(a, b), \T(b) \]
\end{itemize}
We will need to count the possible worlds of this gadget and of
subsequent gadgets, because these quantities will be important in the
reduction to understand the link between the independent-set pairs
of~$G$ and the subinstances of our instances of~$\WUR(Q_1, \phi)$
that satisfy the query.

To this end, we denote by $\lambdaleft$ the
total probability of the possible worlds of the $(a,b)$-gadget that
violate~$Q_1$ when we fix the fact $\R(a)$ to be
present.  We easily compute: $\lambdaleft = 1 - \sigma\tau$.  Similarly, we
denote by~$\lambdaright$ the probability of violating~$Q_1$
when we fix the fact $\T(b)$ to be
present. We have: $\lambdaright = 1 - \rho\sigma$.
In summary, we have the
following notation that we use later in the proof.
\begin{equation}\label{eq:lambdas}
  \lambdaleft = 1-\sigma\tau \quad\quad
  \lambdaright = 1-\rho\sigma \quad\quad
\end{equation}

\begin{itemize}
\item The \underline{\emph{$(a,b,c,d)$-gadget}} is an instance with
  elements $a$, $b$, $c$, and $d$, and the following facts:
    \[
  \R(a), \S(a, b), \T(b), \S(c, b), \R(c), \S(c, d), \T(d)
\]
We illustrate the gadget below, where every vertex represents a domain element,
every edge represents a pair of elements occurring in a fact, and unary and
    binary facts are simply written as relation names, respectively above their element
    and above their edge: 
\begin{center}
\begin{tikzpicture}[xscale=1,yscale=2]
  \node (a) at (0, 0) {$a$};
  \node (b) at (2, 0) {$b$};
  \node (c) at (4, 0) {$c$};
  \node (d) at (6, 0) {$d$};
  \draw (a) edge[->] node[above] {$\S$} (b);
  \draw (c) edge[->] node[above] {$\S$} (b);
  \draw (c) edge[->] node[above] {$\S$} (d);
  \node (aa) at (0, .25) {$\R$};
  \node (bb) at (2, .25) {$\T$};
  \node (cc) at (4, .25) {$\R$};
  \node (dd) at (6, .25) {$\T$};
\end{tikzpicture}
\end{center}

\end{itemize}

We use the following notation for the $(a,b,c,d)$-gadget.
\begin{itemize}
\item $\gamma$ is the total probability of the possible worlds of the gadget that
  violate~$Q_1$ where we fix the facts $\R(a)$ and $\T(d)$
  to be present.
\item $\deltaleft$ is the total probability of possible worlds that
  violate~$Q_1$ when we fix $\R(a)$ to be present and $\T(d)$ to be absent.
\item $\deltaright$ is, symmetrically, the total probability of possible worlds
  that violate~$Q_1$ when we fix $\R(a)$ to be absent and $\T(d)$ to be present.
\item $\deltacut$ is the total probability of possible worlds that
  violate~$Q_1$ when we fix $\R(a)$ and $\T(d)$ to be absent.
 \end{itemize}
 We will study the quantities $\gamma$, $\deltaleft$, $\deltaright$,
 $\deltacut$ in two lemmas in Section~\ref{sec:invertible}.

\paragraph{Defining the reduction.}
Having defined the various gadgets that we will use, let us describe
the instances that we construct from our input bipartite graph
$G = (R \cup T, S)$.    The vertices of~$G$ are the elements
of~$R$ and~$T$, and its edges are the pairs in~$S$.
We write $m \colonequals |S|$, the number of edges of~$G$.

Fix $M \colonequals (\card{R}+1)\times(\card{T}+1)\times(|S|+1)^3$, the
number of instances to which we will reduce. Let us define positive integer
values $B$ and
$B'$ in a somewhat technical way. Specifically, we want 
$B$ to be large enough so that:
\begin{equation}
  \label{eqn:bconstr}
  B > 2m \frac{|\log \gamma| +
  |\log\deltaleft| + |\log \deltaright| + |\log\deltacut|}{
  |\log \lambdaleft|}
\end{equation}
And we then want $B'$ to be large enough so that:
\begin{equation}
  \label{eqn:bpconstr}
  B' > 2 m \frac{|\log \gamma| +
  |\log\deltaleft| + |\log \deltaright| + |\log\deltacut| + 2 B
  |\log\lambdaleft|}{
    |\log \lambdaright|}
\end{equation}
These bounds on~$B$ and~$B'$ will be used later in the proof in a rounding argument,
i.e., arguing that from a value of the form $B' x + By + z$ where $x$, $y$, and
$z$ are bounded, we can recover the separate values $x$ and $y$ and $z$.
Let us explain why this is possible, and why this can be achieved with values that remain polynomial in the input. For this, notice
that $0 < \gamma,\deltaleft,\deltaright,\deltacut < 1$, by definition and
because $0 < \rho, \tau < 1$. Thus, the absolute values of the logarithms of these quantities
are positive numbers, which are constants, i.e., they were fixed with $\phi$ and
do not depend on the instance. This is why we can pick positive integers $B$ and $B'$ to satisfy
these conditions, and the \emph{values} of~$B$ and~$B'$ are 
polynomial in~$m$, so polynomial in the size of the input~$G$.

Now, for each $0 \leq p < (|S|+1)^3$,
$0 \leq \phi \leq \card{R}$, $0 \leq \psi \leq \card{T}$,
we construct the instance $D_{p,\phi,\psi}$ on the schema of~$Q_1$ (i.e.,
two unary relations $\R$ and $\T$ and one binary relation $\S$), as follows.
\begin{itemize}
\item For each vertex $u \in R$ of~$G$, create:
\begin{itemize}
  \item  the fact $\R(u)$;
\item $\phi$ copies of the $(u,*)$-gadget (using fresh elements for
  $b$, as denoted by $*$).
\end{itemize}
\item Similarly, for each vertex $w \in T$ of~$G$ create:
  \begin{itemize}
  \item the fact $\T(w)$;
  \item $\psi$ copies of the $(*,w)$-gadget.
\end{itemize}
\item For every edge $(u, w) \in S$ of~$G$, create:
    \begin{itemize}
    \item $p$ copies of the $(u,*,*,w)$-gadget connecting $u$ and $w$
      (using fresh elements for $b$ and $c$ in each copy);
    \item $B$ copies of the $(u,*)$-gadget
    \item $B'$ copies of the $(*,w)$-gadget
   \end{itemize}
\end{itemize}
Notice that the reduction, i.e., the construction of the instances $D_{p,\phi,
\psi}$ for each $0 \leq p < (|S|+1)^3$,
$0 \leq \phi \leq \card{R}$, $0 \leq \psi \leq \card{T}$ from the input
bipartite graph~$G$, is in polynomial time. Indeed, the number of instances that
we build is polynomial in~$G$, i.e., it is $(|S|+1)^3 \times (\card{R}+1) \times
(\card{T}+1)$. Further, for each instance, for each vertex and edge of~$G$, we
create copies of gadgets of constant size, and the number of copies is $p$, $\phi$,
$\psi$, $B$, or~$B'$: these are polynomial in~$G$, as we pointed out below
Equation~\ref{eqn:bpconstr}.

Further, observe that the construction of~$D_{p,\phi,\psi}$ is
designed to ensure that any match of the query $Q_1$ on a
possible world of~$D_{p,\phi,\psi}$ will always be contained in the
facts of one of the gadgets (plus the facts $\R(u)$ and $\T(w)$).
This means that we can
determine whether the query is true in the possible worlds simply by
looking separately at the facts of each gadget (and at the facts on
the~$u$ and~$w$).

Now, coming back to our reduction, for each choice of $p$, $\phi$, and
$\psi$, we denote
by~$\Pi_{p,\phi,\psi}$ the 
total probability of subinstances of~$D_{p,\phi,\psi}$ that
violate~$Q_1$. Each of these values can be computed in
polynomial time using our oracle for~$\WUR(Q_1, \phi)$: for $0 \leq p < (|S|+1)^3$,
$0 \leq \phi \leq \card{R}$ and $0 \leq \psi \leq \card{T}$, we
build
$D_{p,\phi,\psi}$, call the oracle to obtain the total probability of
instances that satisfy~$Q_1$, and define $\Pi_{p,\phi,\psi}$ as one minus that
number.

Hence, in our reduction, given the input bipartite graph $G$, we have
constructed the instances $D_{p,\phi,\psi}$ and used our oracle
to compute the total probability $\Pi_{p,\phi,\psi}$ of subinstances of each $D_{p,\phi,\psi}$ that
violate~$Q_1$, for each $p,\phi,\psi$, and this process is in
PTIME.
We will show in the sequel how these probabilities can be used to recover the
answer to our original problem
on~$G = (R \cup T, S)$, that is, the
number~$P$ of independent-set pairs of~$G$.

\section{Obtaining the Equation System}
\label{sec:eq-system}

We now move to the second step of our reduction and explain how the
number~$P$ of independent-set pairs of~$G$ is related to the oracle answers~$\Pi_{p,\phi,\psi}$ by a
linear equation system.
To define the linear equation system, it will be helpful to introduce some parameters
about subsets of vertices of the bipartite graph~$G$.
For any
$R' \subseteq R$ and $T' \subseteq T$, we write the following:
\begin{itemize}
\item $\ccc(R', T')$ denotes the number of edges of~$S$ that are
  \emph{contained} in~$R' \times T'$, that is, they have both
  endpoints in $R' \cup T'$. Formally, 
    \[\ccc(R', T') \colonequals\card{(R' \times T') \cap S}.\]
\item $\dd(R', T')$ denotes the number of edges of~$S$ that are
  \emph{dangling from~$R'$}, that is, they have one endpoint in~$R'$
  and the other in $T \setminus T'$. Formally,
    \[\dd(R', T')
    \colonequals \card{ (R' \times (T \setminus T'))\cap S}.\]
\item $\dd'(R', T')$ denotes the number of edges of~$S$ that are
  \emph{dangling from~$T'$}, that is, they have one endpoint in~$R
  \setminus R'$ and the other in $T'$. Formally, 
    \[\dd'(R', T')
    \colonequals \card{((R \setminus R') \times T')\cap S}.\]
\item $\ee(R', T')$ denotes the number of edges of~$S$ that are
  \emph{excluded} from $R' \cup T'$, that is, they have no endpoint in
  $R' \cup T'$. Formally, 
    \[\ee(R', T') \colonequals \card{S \setminus
    (R' \times T')}.\]
\end{itemize}
Clearly, for every $R'$ and $T'$, each edge of~$S$ is either contained
in $R' \times T'$, dangling from~$R'$, dangling from~$T'$, or excluded
from $R' \cup T'$. Hence, 
\[\ccc(R', T') + \dd(R', T') +
\dd'(R', T') + \ee(R', T') = m\,.\]

Observe that a pair $(R',T')$ is an independent-set pair of $G$ iff
$\ccc(R', T') = 0$. Thus, given the input~$G$ to
the reduction, our goal is to compute the following quantity:
\begin{equation}
  \label{eq:defN}
  P = |\{(R', T') \mid R' \subseteq R,\,\,  T' \subseteq T,\,\, \ccc(R', T') = 0\}|
  = \sum_{\substack{R' \subseteq R,~T' \subseteq T,\\ \ccc(R', T') = 0}} \hskip-1.5em 1
\end{equation}

Let us now define the variables of the equation system for the input
graph $G$. We will use these variables to express~$P$, and we will be
able to recover their values from the values~$\Pi_{p,\phi,\psi}$.

\paragraph{Picking variables.}
Our goal is to construct a linear equation system relating the
quantity that we wish to compute, namely~$P$, and the quantities
provided by our oracle, namely $\Pi_{p,\phi,\psi}$ for $0 \leq p < (|S|+1)^3$,
$0 \leq \phi \leq \card{R}$, and $0 \leq \psi \leq \card{T}$. Instead
of using~$P$ directly, we will construct a system
connecting~$\Pi_{p,\phi,\psi}$ to quantities on~$G$ that we now define
and that will allow us to recover~$P$. We call these quantities
\emph{variables} because they are unknown and our goal in the
reduction is to compute them from the~$\Pi_{p,\phi,\psi}$ to
recover~$P$.

Let us introduce, for each $0 \leq i \leq \card{R}$, for each
$0 \leq j \leq \card{T}$, for each $c, d, d' \in\set{0,\dots,|S|}$, the
variable $X_{i,j,c,d,d'}$, that stands for the number of pairs
$(R', T')$ with $\card{R'} = i$, with $\card{T'}=j$, and with $\ccc$-
and $\dd$- and $\dd'$-values exactly as defined earlier in this
section.  (We do not need $e$ as a parameter here because it is
determined from $c$, $d$, and $d'$.) Formally:
\begin{align*}
  X_{i,j,c,d,d'}  \colonequals |\{(R', T') \mid & R' \subseteq R\,,\, T' \subseteq T\,,\,
    \card{R'} = i\,,\, \card{T'} = j, \\
    &  \ccc(R', T') = c\,,\, \dd(R', T') = d\,,\, \dd'(R', T') = d'\}|
\end{align*}
Recall that our fixed query is $Q_1$. 

To simplify the equations that will follow, let us define, for all $i, j, c, d, d'$, other variables,
which are the ones that we will actually use in the equation system:
\[
  Y_{i,j,c,d,d'} \colonequals \rho^i \times (1-\rho)^{\card{R}-i} \times \tau^j
  \times (1-\tau)^{\card{T}-j} \times X_{i,j,c,d,d'}
\]
The reason why we use these slightly more complicated variables is
because they will simplify the equations later.

\paragraph{Getting our answer from the variables.}  Let us now
explain why we can compute our desired value~$P$ (the number of independent-set pairs
of~$G$) from the variables $Y_{i,j,c,d,d'}$.
Refer back to
Equation~\eqref{eq:defN}, and let us split this sum according to the
values of the parameters $i = \card{R'}$, $j = \card{T'}$, and
$\dd(R', T')$, $\dd'(R', T')$. Using our variables $X_{i,j,c,d,d'}$, this gives:
\[
  P = \sum_{0 \leq i \leq \card{R}} \,\sum_{0 \leq j \leq \card{T}} \,\sum_{0 \leq
  d,d' \leq m}
  X_{i,j,0,d,d'}
\]
We can insert the variables $Y_{i,j,c,d,d'}$ instead of $X_{i,j,c,d,d'}$ in the
above, obtaining:
\begin{equation}
  \label{eq:writep}
  P = 
  \sum_{0 \leq i \leq \card{R}}\,
  \sum_{0 \leq j \leq \card{T}}\,
  \sum_{0 \leq d,d' \leq m}
  \frac{Y_{i,j,0,d,d'} }{\rho^i \times (1-\rho)^{\card{R}-i} \times \tau^j
  \times (1-\tau)^{\card{T}-j}}
\end{equation}
This equation justifies that, to compute the quantity $P$ that we are
interested in, it suffices to compute the value of the variables
$Y_{i,j,0,d,d'}$ for all $0 \leq i \leq \card{R}$,
$0 \leq j \leq \card{T}$, and $0 \leq d,d' \leq m$. If we can compute
all $Y_{i,j,0,d,d'}$ in polynomial time, then we can use the equation
above to compute $P$ in polynomial time, completing the reduction.

\paragraph{Designing the equation system.}  We will now design a
linear equation system that connects the quantities $\Pi_{p,\phi,\psi}$
for $p,\phi,\psi$ computed by our oracle to the quantities $Y_{i,j,c,d,d'}$ for all
$0 \leq i \leq \card{R}$, $0 \leq j \leq \card{T}$,
$0 \leq c, d, d' \leq m$ that we wish to compute.  To do so, write the
vector $\vec{\Pi} = (\Pi_{0,0,0}, \ldots, \Pi_{(|S|+1)^3-1,|R|,|T|})$ and
the vector
$\vec{Y} = (Y_{0,0,0,0,0}, \ldots, Y_{\card{R},\card{T},m,m,m})$ in
some order.  We will describe an $M$-by-$M$ matrix $A$ so that we
have the equation $\vec{\Pi} = A \vec{Y}$.  We will later justify that
the matrix $A$ is invertible, so that we can compute $\vec{Y}$ from
$\vec{\Pi}$ and conclude the proof. So, it is left to define $A$, which
we do in the remainder of this section, and to prove that $A$ is
invertible, which we do in the next section.

To define the matrix~$A$, let us consider arbitrary subsets
$R'\subseteq R$ and $T' \subseteq T$, and an arbitrary
$0 \leq p < (|S|+1)^3$. Let us denote by $\D_{p,\phi,\psi}(R', T')$
the set of subinstances of~$D_{p,\phi,\psi}$ where $R'$ is the set of vertices
of $R$ whose $\R$-fact is kept, and where
$T'$ is the set of vertices of~$T$ whose $\T$-fact is kept.
It is clear that the
$\D_{p,\phi,\psi}(R', T')$ form a partition of the subinstances
of~$D_{p,\phi,\psi}$, so that we have the following, where $\Pr$ denotes the
total probability mass:
\begin{equation}\label{eq:Np}
  \Pi_{p,\phi,\psi} = \sum_{R' \subseteq R, T' \subseteq T} \Pr\left(\{I' \in \D_{p,\phi,\psi}(R', T') \mid I'
    \not\models Q_1\}\right)
\end{equation}

Let us now study the number in the above sum for each~$R'$ and $T'$, that is,
the total probability of the
number of instances in~$\D_{p,\phi,\psi}(R', T')$ that violate the query. We can show the
following by performing some accounting over all gadgets in the
construction.
Recall the numbers $\lambdaleft$,
$\lambdaright$, $\deltaleft$, $\deltaright$ and
$\deltacut$ defined in Section~\ref{sec:reduction}. 
\begin{clm}
  \label{clm:countmodels}
For any $0 \leq p <  (|S|+1)^3$, for any choice of~$R'$ and~$T'$,
  writing $i \colonequals \card{R'}$, $j \colonequals \card{T'}$,
  $c \colonequals \ccc(R', T')$, $d \colonequals \dd(R',T')$, $d' \colonequals \dd'(R', T')$,
  $e \colonequals \ee(R',T') = |S| - c  - d - d'$, we have that $\Pr\left(\{I' \in
  \D_{p,\phi,\psi}(R',T') \mid I' \not\models Q_1\}\right)$ is equal to:
  \[
    \rho^i \times (1-\rho)^{\card{R}-i} \times \tau^j \times (1-\tau)^{\card{T}-j} 
    \times (\lambdaleft^i)^\phi \times (\lambdaright^j)^\psi
    \times \alpha(c, d, d')^p
\]
  Where $\alpha(c, d, d')$ is defined as the following quantity:
  \begin{equation}\label{eq:alpha}
  \alpha(c, d, d')\,\colonequals\, \gamma^c \times \deltaleft^d \times
  \deltaright^{d'} \times \deltacut^e
  \times \lambdaleft^{B(c+d)}
  \times \lambdaright^{B'(c+d')}
\end{equation}
\end{clm}

\begin{proof}
  To show this, recall that we can determine whether a possible world
  in~$\D_{p,\phi,\psi}(R',T')$ satisfies the query simply by looking
  at each gadget (and at the facts on the elements $u$ and $w$ in the
  construction of~$D_{p,\phi,\psi}$), as every match of the query can
  use facts from only a single gadget (and possibly the shared facts
  of the $u$ and $w$).
\begin{itemize}
\item We have no choice on the $\R$-facts of $R'$ (we must keep them) and on the
  $\R$-facts of $R \setminus R'$ (we must discard them), inducing a probability of $\rho^{|R'|} \times (1-\rho)^{|R|-|R'|}$.
\item We have no choice on the $\T$-facts of $T'$ (same reasoning), inducing a probability of $\tau^{|T'|} \times (1-\tau)^{|T|-|T'|}$.
\item For each $u \in R'$, we have a probability of $\lambdaleft^{\phi}$
    for the $(u,*)$-gadgets of violating the query. This is true, since
      $\lambdaleft$ is the total probability that a
      $(u,*)$-gadget violates query~$Q_1$ when we fix the
      $\R$-fact on~$u$ to be present, and we consider $\phi$
      copies of the $(u,*)$-gadget.
    \item For each $u \in R \setminus R'$, the $(u,*)$-gadgets cannot be part of
      a query match.
  \item For each $w \in T'$, we have a probability of $\lambdaright^{\psi}$
    that the $(*,w)$-gadgets all violate the query (same reasoning).
   
  \item For each edge $e = (u, w) \in S$ (and using the same reasoning
    as above):
    \begin{itemize}
      \item If $e$ is contained in $R'\times T'$:
    \begin{itemize}
    \item For the $(u,*)$-gadgets, we have
      a probability of $\lambdaleft^{B \times p}$ of violating the query. 
    \item For the $(*,w)$-gadgets, we have a probability of $\lambdaright^{B' \times p}$.
    \item For the $(u,*,*,w)$ gadgets connecting $u$ and
      $w$, we have a probability of $\gamma^p$.
    \end{itemize}
  \item If $e$ is dangling from $R'$:
    \begin{itemize}
    \item For the $(u,*)$-gadgets, we have a probability of $\lambdaleft^{B' \times p}$.
    \item For the $(*,w)$-gadgets, there can be no query match, i.e., a probability of~$1$.
    \item For the $(u,*,*,w)$ gadget connecting $u$ and $w$, we have
      a probability of $\deltaleft^p$.
    \end{itemize}
      \item If $e$ is dangling from~$T'$:
    \begin{itemize}
    \item For the $(u,*)$-gadgets, there can be no query match
    \item For the $(*,w)$-gadgets, we have a probability of $\lambdaright^{B' \times p}$.
    \item For the $(u,*,*,w)$ gadgets connecting $u$ and $w$, we have a probability of $\deltaright^p$.
    \end{itemize}
      \item If $e$ is excluded from $R'\cup T'$:
    \begin{itemize}
    \item For the $(u,*)$-gadgets, there can be no query match.
    \item For the $(*,w)$-gadgets, there can be no query match.
    \item For the $(u,*,*,w)$ gadgets connecting $u$ and $w$, we have
      a probability of $\deltacut^p$.
    \end{itemize}
    \end{itemize}
\end{itemize}
Therefore, with $i = \card{R'}$, $j = \card{T'}$, $c = \ccc(R', T')$, $d
= \dd(R',T')$, $d' = \dd'(R', T')$, $e = \ee(R',T')$, we have:
\begin{align*}
  &  \Pr\left(I' \in
  \D_{p,\phi,\psi}(R',T') \mid I' \not\models Q_1\right) =
  \rho^i \times (1-\rho)^{\card{R}-i} \times \tau^j \times (1-\tau)^{\card{T}-j}
  \\
  & \quad\times \gamma^{cp} \times \deltaleft^{dp} \times \deltaright^{d'p} \times \deltacut^{ep}\times \lambdaleft^{(B(c+d))p}
  \times \lambdaright^{(B'(c+d'))p} \times \lambdaleft^{i\phi} \times \lambdaright^{j\psi}
\end{align*}
This leads directly to the claimed result.
\end{proof}

Let us use the value of Claim~\ref{clm:countmodels} in
Equation~\eqref{eq:Np}. Note that this value only depends on the
cardinalities of~$R'$ and~$T'$ and the values
of~$\ccc, \dd, \dd', \ee$, but not on the specific choice of~$R'$
and~$T'$. Thus, splitting the sum accordingly, we can obtain the
following:

\begin{clm}
  \label{clm:matrixeqn}
  For any $0 \leq p <  (|S|+1)^3$, we have that:
\[
  \Pi_{p,\phi,\psi} = \sum_{\substack{0 \leq i \leq \card{R}\\0 \leq j \leq \card{T}\\0\leq c, d, d'
  \leq m}}
  Y_{i,j,c,d,d'} \times (\lambdaleft^i)^\phi \times (\lambdaright^j)^\psi \times
  \alpha(c, d, d')^p \,.
\]
\end{clm}
\begin{proof}
  Substituting the equality from Claim~\ref{clm:countmodels} into
  Equation~\eqref{eq:Np}, and splitting the sum,
  we get:
  \[
  \Pi_{p,\phi,\psi} = \hspace{-1em}\sum_{\substack{0 \leq i \leq \card{R}\\0 \leq j \leq \card{T}\\0\leq c, d, d'
  \leq m}} 
  \sum_{\substack{R' \subseteq R\\T' \subseteq T\\\card{R'} = i, \card{T'} =
    j\\\ccc(R', T') = c\\\dd(R', T') = d, \\\dd'(R', T') = d'}}\hspace{-1em}
  \rho^i \times (1-\rho)^{\card{R}-i} 
    \times
    \tau^j \times (1-\tau)^{\card{T}-j}
\times (\lambdaleft^{i})^\phi
  \times (\lambdaright^{j})^\psi
    \times \alpha(c, d, d')^p
  \]
The inner sum does not depend on~$R'$ and~$T'$, so let us introduce the variables
$X_{i,j,c,d,d'}$:
  \[
  \Pi_{p,\phi,\psi} = \hspace{-1em}\sum_{\substack{0 \leq i \leq \card{R}\\ 0 \leq j \leq \card{T}\\0\leq c, d, d'
  \leq m}} \hspace{-1em} 
  X_{i,j,c,d,d'} \times
  \rho^i \times (1-\rho)^{\card{R}-i}
  \times 
  \tau^j \times  (1-\tau)^{\card{T}-j} 
  \times (\lambdaleft^{i})^\phi
  \times (\lambdaright^{j})^\psi 
  \times \alpha(c, d, d')^p
\]
Note that we can now use the variables $Y_{i,j,c,d,d'}$ to eliminate the
  remaining factors, obtaining the claimed equality. Note that this simplification is the reason why we
  introduced the variables $Y_{i,j,c,d,d'}$, to make the equality more
  convenient to work with.
\end{proof}

The equation of Claim~\ref{clm:matrixeqn} can be expressed as a matrix equation $\vec{\Pi} = A' \vec{Y}$, with
$A'$ the matrix defined by \begin{equation}\label{eq:matrix}
  A'_{(\phi,\psi,p),(i,j,c,d,d')} \colonequals (\lambdaleft^i)^\phi \times
(\lambdaleft^j)^\psi \times \alpha(c, d, d')^p\,.\end{equation}
The matrix~$A'$ relates the vector $\vec{\Pi}$ computed from our oracle
calls and the variables $\vec{Y}$ that we wish to determine to solve
our problem on the graph $G$.  It remains to show that $A'$ is an
invertible matrix, so that we can compute its inverse $(A')^{-1}$ in
polynomial time, use it to recover $\vec{Y}$ from $\vec{\Pi}$, and from
there recover~$P$ via Equation~\eqref{eq:writep}, concluding the
reduction.

To study the invertibility of the matrix~$A'$, we will notice that it can be
expressed as the \emph{Kronecker product} of three matrices that we will show to
be invertible. Recall that the \emph{Kronecker product} of a
$\kappa \times \kappa$ matrix $C$ and of a $\nu \times \nu$ matrix
$C'$ is the $(\kappa\nu) \times (\kappa\nu)$-matrix with the following
blockwise definition:
\[
C \otimes C' = \begin{bmatrix}
  C_{1,1} C' & C_{1,2} C'& \dots  & C_{1,\kappa} C'\\
    C_{2,1} C' & C_{2,2} C' & \dots  & C_{2,\kappa} \\
    \vdots & \vdots & \ddots & \vdots \\
    C_{\kappa,1} C' & C_{\kappa,2} C' & \dots  &
    C_{\kappa,\kappa} C
\end{bmatrix}
\]
We refer the reader to literature such as Henderson, Pukelsheim, and
Searle~\cite{henderson1983history} for the details and history of the
Kronecker product.

Now, following the definition of~$A'$ in Equation~\eqref{eq:matrix}, and numbering
the rows lexicographically by $(\phi, \psi, p)$ with 
$0 \leq \phi \leq \card{R}$, $0 \leq \psi \leq \card{T}$, and
$0 \leq p < (|S|+1)^3$, and numbering the columns lexicographically by $(i, j,
(c,d,d'))$ with 
$0 \leq i \leq \card{R}$, $0 \leq j \leq \card{T}$, and
$c, d, d' \in\set{0,\dots,|S|}$, we see that the matrix $A'$ 
is by definition the Kronecker product of three matrices:
\begin{enumerate}
  \item The $(\card{R}+1)\times(\card{R}+1)$ matrix $W$ whose cell $(\phi,i)$
    contains $(\lambdaleft^i)^\phi$;
  \item The $(\card{T}+1)\times(\card{T}+1)$ matrix $W'$ whose cell $(\psi,j)$
    contains $(\lambdaright^j)^\psi$;
\item The $(\card{S}+1)^3\times(\card{S}+1)^3$ matrix $A$ whose cell $p,(c,d,d')$ contains
  $\alpha(c,d,d')^p$.
\end{enumerate}

We know that the Kronecker product of invertible matrices is
invertible, so to show that $A'$ is invertible, it suffices to show
that $W$, $W'$, and $A$ are invertible. Now, the matrix $W$ is clearly
a (transpose of a) Vandermonde
matrix,\footnote{Recall that an $m\times m$ matrix is a
    \e{Vandermonde matrix} if there are $m$ numbers $x_1$,\dots,$x_m$
    such that each cell $(k,\ell)$ is $x_k^{\ell-1}$. It is known that such a
    matrix is invertible
    if and only if $x_k\neq x_{k'}$ for $k\neq k'$.}  and it is invertible: we
have $0 < \tau < 1$ from which the definition of $\lambdaleft$ implies
$0 < \lambdaleft < 1$, so the function mapping an integer $i$ to
$\lambdaleft^i$ is injective.  The same reasoning shows that $W'$ is
invertible.  Hence, it suffices to study if~$A$ is invertible. This
matrix is clearly also a Vandermonde matrix, so the only remaining
point to show that $A'$ is invertible is to show that the coefficients
$\alpha(c, d, d')$ of~$A$ are different. We do this in the next
section.

\section{Showing that the Matrix is Invertible}
\label{sec:invertible}

This section presents the third step of the proof of Theorem~\ref{thm:Qrst-hard}
and concludes. Specifically, we show the following:

\begin{clm}
  \label{clm:injective}
  For all $0 \leq c_1, c_2, d_1, d_2, d_1', d_2' \leq m$, if 
  $(c_1, d_1, d_1') \neq (c_2,d_2, d'_2)$, then we have
  $\alpha(c_1, d_1, d_1') \neq \alpha(c_2, d_2, d_2')$, where $\alpha$ is as defined in Claim~\ref{clm:countmodels}.
\end{clm}
Claim~\ref{clm:injective} implies that the Vandermonde matrix~$A$ is
invertible, and concludes the definition of the reduction and the
proof of Theorem~\ref{thm:Qrst-hard}.

Let us show the contrapositive of the statement: we take
  $0 \leq c_1, c_2, d_1, d_2, d_1', d_2' \leq m$ such that 
  $\alpha(c_1, d_1, d_1') = \alpha(c_2, d_2, d_2')$, and we must show that $c_1
  = c_2$, $d_1 = d_2$, and $d_1' = d_2'$.

We know that the values of $\alpha$ are positive, so let us inject the
definition of $\alpha$ from Claim~\ref{clm:countmodels} and take the logarithm
of the equality. We obtain the following, where $e_1 \colonequals m - c_1 - d_1
- d_1'$, and likewise $e_2 \colonequals m - c_2 - d_2
- d_2'$:
\begin{multline*}
  c_1 \log \gamma + d_1 \log \deltaleft +
  d'_1 \log \deltaright + e_1 \log\deltacut
  + B(c_1+d_1) \log\lambdaleft
  + B'(c_1+d_1') \log\lambdaright\\
  = 
  c_2 \log \gamma + d_2 \log \deltaleft +
  d'_2 \log \deltaright + e_2 \log\deltacut
  + B(c_2+d_2) \log\lambdaleft
  + B'(c_2+d_2') \log\lambdaright
\end{multline*}
As $0 < \rho, \tau < 1$, we clearly have $0 < \gamma, \deltaleft,
\deltaright, \deltacut, \lambdaleft, \lambdaright < 1$, and their logarithms are negative quantities.
Let us divide the equation by $B' \log\lambdaright$, to obtain:
\begin{multline*}
  \frac{c_1 \log \gamma + d_1 \log \deltaleft +
  d'_1 \log \deltaright + e_1 \log\deltacut
  + B(c_1+d_1) \log\lambdaleft}{B' \log\lambdaright}
  + c_1+d_1' \\
  \hfill = \frac{c_2 \log \gamma + d_2 \log \deltaleft +
  d'_2 \log \deltaright + e_2 \log\deltacut
  + B(c_2+d_2) \log\lambdaleft}{B' \log\lambdaright}
  + c_2+d_2'
\end{multline*}
Recall from Equation~\eqref{eqn:bpconstr} that our definition of $B'$ makes it
large enough to ensure that $(m (\log \gamma +
\log\deltaleft + \log \deltaright + \log\deltacut) + 2 m B \log\lambdaleft) /
B' \log \lambdaright$ has absolute value $<0.5$, using the triangle inequality.
Thus, we can bound the absolute value of the first term of the left-hand side of the equation by~$0.5$.
The same applies to the first term of the
right-hand side of the equation. By contrast, $c_1 + d_1'$ and $c_2 + d_2'$ are
integers.
Thus, by a rounding argument, we
conclude:
\begin{equation}
  \label{eq:cdp}
  c_1 + d_1' = c_2 + d_2'
\end{equation}

Simplifying away the common term, multiplying back by $B' \log
\lambdaright$, and dividing by $B \log \lambdaleft$, we obtain:
\begin{multline*}
  \frac{c_1 \log \gamma + d_1 \log \deltaleft +
  d'_1 \log \deltaright + e_1 \log\deltacut}{B \log\lambdaleft} +
  c_1 + d_1 \\
  = 
  \frac{c_2 \log \gamma + d_2 \log \deltaleft +
  d'_2 \log \deltaright + e_2 \log\deltacut}{B \log\lambdaleft} +
  c_2 + d_2
\end{multline*}
Recall from Equation~\eqref{eqn:bconstr} that our definition of $B$ makes it large
enough to ensure that $m (\log \gamma +
\log\deltaleft + \log \deltaright + \log\deltacut) /
B \log \lambdaleft$ has absolute value $<.5$. Thus we conclude again by a rounding
argument that:
\begin{equation}
  \label{eq:cd}
  c_1 + d_1 = c_2 + d_2
\end{equation}

Finally, simplifying the common term, multiplying back by $B \log \lambdaleft$,
and exponentiating, we obtain a third equation:
\begin{equation}
  \label{eq:final}
  \gamma^{c_1} \times \deltaleft^{d_1} \times
  \deltaright^{d_1'} \times \deltacut^{e_1}
  = 
  \gamma^{c_2} \times \deltaleft^{d_2} \times
  \deltaright^{d_2'} \times \deltacut^{e_2}
\end{equation}

Let $x = d_1-d_2$ be the remaining degree of freedom. From
Equation~\eqref{eq:cd}, we know that $x = c_2-c_1$. From this and
Equation~\eqref{eq:cdp}, we know that $x = d_1' - d_2'$. Combining these three
definitions of $x$, we get that $x - x + x = (d_1-d_2) - (c_2-c_1) + (d_1'-d_2')$,
so that $x = c_1+d_1+d_1' - c_2-d_2-d_2'$, i.e., $x = e_2 - e_1$.
Thus, we can rewrite
Equation~\eqref{eq:final} to:
\begin{equation}
  \label{eq:final2}
  \left(\frac{
    \deltaleft \times \deltaright
  }{
    \gamma \times \deltacut
  }\right)^x = 1
\end{equation}

Our goal is to show that $x = 0$. If true, then from the definitions of~$x$
above, it implies $c_1 = c_2$, $d_1 = d_2$, and $d_1' = d_2'$, what we want to
show. For this the key claim is to show that the fraction being exponentiated is
not equal to~$1$. Formally:

\begin{lem}
  \label{lem:diseq}
  $\deltaleft \times \deltaright \neq \gamma \times \deltacut$.
\end{lem}

\begin{proof}
  To show this result, we will have to study in detail the quantities 
  $\deltaleft \times \deltaright$ and $\gamma \times \deltacut$. To this end,
  to simplify the presentation, we name some variants of the
  $(a,b,c,d)$-gadget. 
\begin{itemize}
  \item The \underline{\emph{$(a,b,c,d)$-full-gadget}} is like the
    $(a,b,c,d)$-gadget, but we fix the facts $\R(a)$ and $\T(d)$ 
    to be present, so that $\gamma$ is the total probability of possible
    worlds of such a gadget which violate $Q_1$.
\item The \underline{\emph{$(a,b,c,d)$-left-gadget}} is like the
  $(a,b,c,d)$-gadget, but we fix the fact $\R(a)$ to be present and the fact
    $\T(d)$ to be absent, so that $\deltaleft$ is the
    total probability of possible worlds of such a gadget which violate $Q_1$.
  \item The \underline{\emph{$(a,b,c,d)$-right-gadget}} is like the
    $(a,b,c,d)$-gadget, but we fix the fact $\T(d)$ to be present and the fact
    $\R(a)$ to be absent, so $\deltaright$ is the total probability  of
    possible worlds of such a gadget which violate $Q_1$.

  \item The \underline{\emph{$(a,b,c,d)$-trimmed-gadget}} is like the
    $(a,b,c,d)$-gadget but where we fix the facts $\R(a)$ and $\T(d)$ to be
    absent, so $\deltacut$ is the total probability of possible
    worlds of such a gadget which violate $Q_1$.
\end{itemize}
Let us consider, on the one hand, a pair of an $(a,b,c,d)$-left-gadget
and of a $(a',b',c',d')$-right-gadget (giving distinct names to each
vertex).
  As previously we 
  indicate the facts $\R$, $\S$, and $\T$ graphically. 
  Further, we will write $\R$ in the following way:
  \begin{itemize}
  \item as $\underline\R$ to mean that the fact was fixed to be 
    present, that is, as we did on the endpoints when
    defining~$\gamma$;
  \item as $\notR$ to mean that the fact was fixed to be missing (intuitively,
    the vertex cannot be used for a match of the query);
  \item as $\R?$ to mean that we have not yet fixed the fact
  \end{itemize}
  We do the same for the $\S$-facts and $\T$-facts.
  Under these conventions, the left and right gadgets are,
  respectively:
  
  \medskip
  \begin{tikzpicture}[xscale=1,yscale=2]
  \node (a) at (0, 0) {$a$};
    \node (b) at (2, 0) {$b$};
    \node (c) at (4, 0) {$c$};
  \node (d) at (6, 0) {$d$};
  \draw (a) edge[->] node[above] {$\S$?} (b);
  \draw (c) edge[->] node[above] {$\S$?} (b);
  \draw (c) edge[->] node[above] {$\S$?} (d);
  \node (aa) at (0, .25) {$\underline\R$};
    \node (bb) at (2, .25) {$\T$?};
    \node (cc) at (4, .25) {$\R$?};
    \node (dd) at (6, .25) {$\notT$};
  \end{tikzpicture}\hfill
  \begin{tikzpicture}[xscale=1,yscale=2]
  \node (a) at (0, 0) {$a'$};
    \node (b) at (2, 0) {$b'$};
    \node (c) at (4, 0) {$c'$};
  \node (d) at (6, 0) {$d'$};
  \draw (a) edge[->] node[above] {$\S$?} (b);
  \draw (c) edge[->] node[above] {$\S$?} (b);
  \draw (c) edge[->] node[above] {$\S$?} (d);
    \node (aa) at (0, .25) {$\notR$};
    \node (bb) at (2, .25) {$\T$?};
    \node (cc) at (4, .25) {$\R$?};
  \node (dd) at (6, .25) {$\underline\T$};
  \end{tikzpicture}
  \medskip

  And let us consider, on the other hand, a pair of an $(e,f,g,h)$-full-gadget and of an
$(e',f',g',h')$-trimmed-gadget, which we represent in the same way:
  
  \medskip
  \begin{tikzpicture}[xscale=1,yscale=2]
  \node (a) at (0, 0) {$e$};
    \node (b) at (2, 0) {$f$};
    \node (c) at (4, 0) {$g$};
  \node (d) at (6, 0) {$h$};
  \draw (a) edge[->] node[above] {$\S$?} (b);
  \draw (c) edge[->] node[above] {$\S$?} (b);
  \draw (c) edge[->] node[above] {$\S$?} (d);
  \node (aa) at (0, .25) {$\underline\R$};
    \node (bb) at (2, .25) {$\T$?};
    \node (cc) at (4, .25) {$\R$?};
  \node (dd) at (6, .25) {$\underline\T$};
  \end{tikzpicture}\hfill
  \begin{tikzpicture}[xscale=1,yscale=2]
  \node (a) at (0, 0) {$e'$};
    \node (b) at (2, 0) {$f'$};
    \node (c) at (4, 0) {$g'$};
  \node (d) at (6, 0) {$h'$};
  \draw (a) edge[->] node[above] {$\S$?} (b);
  \draw (c) edge[->] node[above] {$\S$?} (b);
  \draw (c) edge[->] node[above] {$\S$?} (d);
    \node (aa) at (0, .25) {$\notR$};
    \node (bb) at (2, .25) {$\T$?};
    \node (cc) at (4, .25) {$\R$?};
    \node (dd) at (6, .25) {$\notT$};
  \end{tikzpicture}
  \medskip

  Let us compute the difference $\Delta \colonequals \deltaleft \deltaright -
  \deltacut \gamma$, and show that it is non-zero.

  The first term, $\deltaleft \deltaright$, is the total probability of possible worlds of
  the first figure that violate the query. It is clearly unchanged when fixing
  the fact $\S(c, d)$ to be missing, because this fact cannot participate to a
  match of the query~$Q_1$ because $\T(d)$ is missing, and likewise when fixing
  $\S(a', b')$ to be missing because $\R(a')$ is missing. 
Likewise, for $\gamma
  \deltacut$,
  we can fix $\S(e', f')$ and $\S(g', h')$ to be missing.

  Let us write the remaining parts of the gadgets, where we remove the edges
  that we have fixed and can no longer intervene in a match of the query:

  \medskip
  \begin{tikzpicture}[xscale=1,yscale=2]
  \node (a) at (0, 0) {$a$};
    \node (b) at (2, 0) {$b$};
    \node (c) at (4, 0) {$c$};
  \draw (a) edge[->] node[above] {$\S$?} (b);
  \draw (c) edge[->] node[above] {$\S$?} (b);
  \node (aa) at (0, .25) {$\underline\R$};
    \node (bb) at (2, .25) {$\T$?};
    \node (cc) at (4, .25) {$\R$?};
  \end{tikzpicture}\hfill
  \begin{tikzpicture}[xscale=1,yscale=2]
    \node (b) at (2, 0) {$b'$};
    \node (c) at (4, 0) {$c'$};
  \node (d) at (6, 0) {$d'$};
  \draw (c) edge[->] node[above] {$\S$?} (b);
  \draw (c) edge[->] node[above] {$\S$?} (d);
    \node (bb) at (2, .25) {$\T$?};
    \node (cc) at (4, .25) {$\R$?};
  \node (dd) at (6, .25) {$\underline\T$};
  \end{tikzpicture}
  \medskip

  \medskip
  \begin{tikzpicture}[xscale=1,yscale=2]
  \node (a) at (0, 0) {$e$};
    \node (b) at (2, 0) {$f$};
    \node (c) at (4, 0) {$g$};
  \node (d) at (6, 0) {$h$};
  \draw (a) edge[->] node[above] {$\S$?} (b);
  \draw (c) edge[->] node[above] {$\S$?} (b);
  \draw (c) edge[->] node[above] {$\S$?} (d);
  \node (aa) at (0, .25) {$\underline\R$};
    \node (bb) at (2, .25) {$\T$?};
    \node (cc) at (4, .25) {$\R$?};
  \node (dd) at (6, .25) {$\underline\T$};
  \end{tikzpicture}\hfill
  \begin{tikzpicture}[xscale=1,yscale=2]
    \node (b) at (2, 0) {$f'$};
    \node (c) at (4, 0) {$g'$};
  \draw (c) edge[->] node[above] {$\S$?} (b);
    \node (bb) at (2, .25) {$\T$?};
    \node (cc) at (4, .25) {$\R$?};
  \end{tikzpicture}
  \medskip

  We can now write:
  \[\deltaleft = (1-\sigma)\deltaleft^- + \deltaleft^+\]
  where $\deltaleft^+$ is the total probability of possible worlds
  violating~$Q_1$ where we keep the fact
  $\S(a,b)$, and $\deltaleft^-$ the total probability of possible worlds
  violating~$Q_1$ where this fact is
  missing.

  Likewise, we write:
  \[\gamma = (1-\sigma)\gamma^- + \gamma^+\]
  where $\gamma^+$ is the total probability of possible worlds violating~$Q_1$ where we keep the fact
  $\S(e,f)$, and $\gamma^-$ the total probability of possible worlds
  violating~$Q_1$ where at least one of these facts is missing.

  Thus, we have:
  \[
    \Delta = (1-\sigma) \Delta^- + \sigma \Delta^+
  \]
  where \[\Delta^- = \deltaleft^- \deltaright - \gamma^-\deltacut
    \qquad\qquad \Delta^+ = \deltaleft^+ \deltaright - \gamma^+\deltacut\,.\]

  But let us consider the status of the gadgets in the choices made in
  $\Delta^-$. They are:

  \medskip
  \begin{tikzpicture}[xscale=1,yscale=2]
  \node (a) at (0, 0) {$a$};
    \node (b) at (2, 0) {$b$};
    \node (c) at (4, 0) {$c$};
  \draw (a) edge[->] node[above] {$\notS$} (b);
  \draw (c) edge[->] node[above] {$\S$?} (b);
  \node (aa) at (0, .25) {$\underline\R$};
    \node (bb) at (2, .25) {$\T$?};
    \node (cc) at (4, .25) {$\R$?};
  \end{tikzpicture}\hfill
  \begin{tikzpicture}[xscale=1,yscale=2]
    \node (b) at (2, 0) {$b'$};
    \node (c) at (4, 0) {$c'$};
  \node (d) at (6, 0) {$d'$};
  \draw (c) edge[->] node[above] {$\S$?} (b);
  \draw (c) edge[->] node[above] {$\S$?} (d);
    \node (bb) at (2, .25) {$\T$?};
    \node (cc) at (4, .25) {$\R$?};
  \node (dd) at (6, .25) {$\underline\T$};
  \end{tikzpicture}
  \medskip

  \medskip
  \begin{tikzpicture}[xscale=1,yscale=2]
  \node (a) at (0, 0) {$e$};
    \node (b) at (2, 0) {$f$};
    \node (c) at (4, 0) {$g$};
  \node (d) at (6, 0) {$h$};
  \draw (a) edge[->] node[above] {$\notS$} (b);
  \draw (c) edge[->] node[above] {$\S$?} (b);
  \draw (c) edge[->] node[above] {$\S$?} (d);
  \node (aa) at (0, .25) {$\underline\R$};
    \node (bb) at (2, .25) {$\T$?};
    \node (cc) at (4, .25) {$\R$?};
  \node (dd) at (6, .25) {$\underline\T$};
  \end{tikzpicture}\hfill
  \begin{tikzpicture}[xscale=1,yscale=2]
    \node (b) at (2, 0) {$f'$};
    \node (c) at (4, 0) {$g'$};
  \draw (c) edge[->] node[above] {$\S$?} (b);
    \node (bb) at (2, .25) {$\T$?};
    \node (cc) at (4, .25) {$\R$?};
  \end{tikzpicture}
  \medskip

  The absence of the fact $\S(a, b)$ means that this edge can no longer
  be part of a match of the query, and the same holds for $(e, f)$, yielding:

  \medskip
  \begin{tikzpicture}[xscale=1,yscale=2]
    \node (b) at (2, 0) {$b$};
    \node (c) at (4, 0) {$c$};
  \draw (c) edge[->] node[above] {$\S$?} (b);
    \node (bb) at (2, .25) {$\T$?};
    \node (cc) at (4, .25) {$\R$?};
  \end{tikzpicture}\hfill
  \begin{tikzpicture}[xscale=1,yscale=2]
    \node (b) at (2, 0) {$b'$};
    \node (c) at (4, 0) {$c'$};
  \node (d) at (6, 0) {$d'$};
  \draw (c) edge[->] node[above] {$\S$?} (b);
  \draw (c) edge[->] node[above] {$\S$?} (d);
    \node (bb) at (2, .25) {$\T$?};
    \node (cc) at (4, .25) {$\R$?};
  \node (dd) at (6, .25) {$\underline\T$};
  \end{tikzpicture}
  \medskip

  \medskip
  \begin{tikzpicture}[xscale=1,yscale=2]
    \node (b) at (2, 0) {$f$};
    \node (c) at (4, 0) {$g$};
  \node (d) at (6, 0) {$h$};
  \draw (c) edge[->] node[above] {$\S$?} (b);
  \draw (c) edge[->] node[above] {$\S$?} (d);
    \node (bb) at (2, .25) {$\T$?};
    \node (cc) at (4, .25) {$\R$?};
  \node (dd) at (6, .25) {$\underline\T$};
  \end{tikzpicture}\hfill
  \begin{tikzpicture}[xscale=1,yscale=2]
    \node (b) at (2, 0) {$f'$};
    \node (c) at (4, 0) {$g'$};
  \draw (c) edge[->] node[above] {$\S$?} (b);
    \node (bb) at (2, .25) {$\T$?};
    \node (cc) at (4, .25) {$\R$?};
  \end{tikzpicture}
  \medskip

  We note that the gadgets of the first line are isomorphic to that of the
  second line, so they have the same
  probability of violating~$Q_1$. Thus, $\Delta^- = 0$, and
  $\Delta = \sigma \Delta^+$.

  We accordingly study the status of the gadgets in $\Delta^+$:

  \medskip
  \begin{tikzpicture}[xscale=1,yscale=2]
  \node (a) at (0, 0) {$a$};
    \node (b) at (2, 0) {$b$};
    \node (c) at (4, 0) {$c$};
  \draw (a) edge[->] node[above] {$\underline\S$} (b);
  \draw (c) edge[->] node[above] {$\S$?} (b);
  \node (aa) at (0, .25) {$\underline\R$};
    \node (bb) at (2, .25) {$\T$?};
    \node (cc) at (4, .25) {$\R$?};
  \end{tikzpicture}\hfill
  \begin{tikzpicture}[xscale=1,yscale=2]
    \node (b) at (2, 0) {$b'$};
    \node (c) at (4, 0) {$c'$};
  \node (d) at (6, 0) {$d'$};
  \draw (c) edge[->] node[above] {$\S$?} (b);
  \draw (c) edge[->] node[above] {$\S$?} (d);
    \node (bb) at (2, .25) {$\T$?};
    \node (cc) at (4, .25) {$\R$?};
  \node (dd) at (6, .25) {$\underline\T$};
  \end{tikzpicture}
  \medskip

  \medskip
  \begin{tikzpicture}[xscale=1,yscale=2]
  \node (a) at (0, 0) {$e$};
    \node (b) at (2, 0) {$f$};
    \node (c) at (4, 0) {$g$};
  \node (d) at (6, 0) {$h$};
  \draw (a) edge[->] node[above] {$\underline\S$} (b);
  \draw (c) edge[->] node[above] {$\S$?} (b);
  \draw (c) edge[->] node[above] {$\S$?} (d);
  \node (aa) at (0, .25) {$\underline\R$};
    \node (bb) at (2, .25) {$\T$?};
    \node (cc) at (4, .25) {$\R$?};
  \node (dd) at (6, .25) {$\underline\T$};
  \end{tikzpicture}\hfill
  \begin{tikzpicture}[xscale=1,yscale=2]
    \node (b) at (2, 0) {$f'$};
    \node (c) at (4, 0) {$g'$};
  \draw (c) edge[->] node[above] {$\S$?} (b);
    \node (bb) at (2, .25) {$\T$?};
    \node (cc) at (4, .25) {$\R$?};
  \end{tikzpicture}
  \medskip

  We write in the same way:
  \[\deltaright = (1-\sigma)\deltaright^- + \deltaright^+\]
  Where $\deltaright^+$ is the total probability of possible worlds
  violating~$Q_1$ where we keep the fact
  $\S(c,d)$, and $\deltaright^-$ is the total probability of possible worlds
  violating~$Q_1$ where at
  least one of these facts is missing.

  We also write:
  \[\gamma^+ = (1-\sigma)\gamma^{+-} + \gamma^{++}\]
  Where $\gamma^{++}$ is the total probability of possible worlds
  violating~$Q_1$ where we keep the fact
  $\S(g,h)$, and $\gamma^{+-}$ is the total probability of possible worlds where at
  least one of these facts is missing.

  Thus, we have:
  \[
    \Delta^+ = (1-\sigma) \Delta^{+-} + \sigma \Delta^{++}
  \]
  where \[\Delta^{+-} = \deltaleft^+ \deltaright^- - \gamma^{+-} \deltacut
  \qquad\qquad \Delta^{++} = \deltaleft^+ \deltaright^+ - \gamma^{++}\deltacut\,.\]

  And let us consider again the choices made in $\Delta^{+-}$:

  \medskip
  \begin{tikzpicture}[xscale=1,yscale=2]
  \node (a) at (0, 0) {$a$};
    \node (b) at (2, 0) {$b$};
    \node (c) at (4, 0) {$c$};
  \draw (a) edge[->] node[above] {$\underline\S$} (b);
  \draw (c) edge[->] node[above] {$\S$?} (b);
  \node (aa) at (0, .25) {$\underline\R$};
    \node (bb) at (2, .25) {$\T$?};
    \node (cc) at (4, .25) {$\R$?};
  \end{tikzpicture}\hfill
  \begin{tikzpicture}[xscale=1,yscale=2]
    \node (b) at (2, 0) {$b'$};
    \node (c) at (4, 0) {$c'$};
  \node (d) at (6, 0) {$d'$};
  \draw (c) edge[->] node[above] {$\S$?} (b);
  \draw (c) edge[->] node[above] {$\notS$} (d);
    \node (bb) at (2, .25) {$\T$?};
    \node (cc) at (4, .25) {$\R$?};
  \node (dd) at (6, .25) {$\underline\T$};
  \end{tikzpicture}
  \medskip

  \medskip
  \begin{tikzpicture}[xscale=1,yscale=2]
  \node (a) at (0, 0) {$e$};
    \node (b) at (2, 0) {$f$};
    \node (c) at (4, 0) {$g$};
  \node (d) at (6, 0) {$h$};
  \draw (a) edge[->] node[above] {$\underline\S$} (b);
  \draw (c) edge[->] node[above] {$\S$?} (b);
  \draw (c) edge[->] node[above] {$\notS$} (d);
  \node (aa) at (0, .25) {$\underline\R$};
    \node (bb) at (2, .25) {$\T$?};
    \node (cc) at (4, .25) {$\R$?};
  \node (dd) at (6, .25) {$\underline\T$};
  \end{tikzpicture}\hfill
  \begin{tikzpicture}[xscale=1,yscale=2]
    \node (b) at (2, 0) {$f'$};
    \node (c) at (4, 0) {$g'$};
  \draw (c) edge[->] node[above] {$\S$?} (b);
    \node (bb) at (2, .25) {$\T$?};
    \node (cc) at (4, .25) {$\R$?};
  \end{tikzpicture}
  \medskip

  Again, we eliminate the fixed parts of the gadget that cannot contribute to a query
  match, and obtain:

  \medskip
  \begin{tikzpicture}[xscale=1,yscale=2]
  \node (a) at (0, 0) {$a$};
    \node (b) at (2, 0) {$b$};
    \node (c) at (4, 0) {$c$};
  \draw (a) edge[->] node[above] {$\underline\S$} (b);
  \draw (c) edge[->] node[above] {$\S$?} (b);
  \node (aa) at (0, .25) {$\underline\R$};
    \node (bb) at (2, .25) {$\T$?};
    \node (cc) at (4, .25) {$\R$?};
  \end{tikzpicture}\hfill
  \begin{tikzpicture}[xscale=1,yscale=2]
    \node (b) at (2, 0) {$b'$};
    \node (c) at (4, 0) {$c'$};
  \draw (c) edge[->] node[above] {$\S$?} (b);
    \node (bb) at (2, .25) {$\T$?};
    \node (cc) at (4, .25) {$\R$?};
  \end{tikzpicture}
  \medskip

  \medskip
  \begin{tikzpicture}[xscale=1,yscale=2]
  \node (a) at (0, 0) {$e$};
    \node (b) at (2, 0) {$f$};
    \node (c) at (4, 0) {$g$};
  \draw (a) edge[->] node[above] {$\underline\S$} (b);
  \draw (c) edge[->] node[above] {$\S$?} (b);
  \node (aa) at (0, .25) {$\underline\R$};
    \node (bb) at (2, .25) {$\T$?};
    \node (cc) at (4, .25) {$\R$?};
  \end{tikzpicture}\hfill
  \begin{tikzpicture}[xscale=1,yscale=2]
    \node (b) at (2, 0) {$f'$};
    \node (c) at (4, 0) {$g'$};
  \draw (c) edge[->] node[above] {$\S$?} (b);
    \node (bb) at (2, .25) {$\T$?};
    \node (cc) at (4, .25) {$\R$?};
  \end{tikzpicture}
  \medskip

  And again these gadgets are isomorphic, so $\Delta^{+-} = 0$ and $\Delta^+ =
  \sigma \Delta^{++}$.

  The gadgets in $\Delta^{++}$ are:
  
  \medskip
  \begin{tikzpicture}[xscale=1,yscale=2]
  \node (a) at (0, 0) {$a$};
    \node (b) at (2, 0) {$b$};
    \node (c) at (4, 0) {$c$};
  \draw (a) edge[->] node[above] {$\underline\S$} (b);
  \draw (c) edge[->] node[above] {$\S$?} (b);
  \node (aa) at (0, .25) {$\underline\R$};
    \node (bb) at (2, .25) {$\T$?};
    \node (cc) at (4, .25) {$\R$?};
  \end{tikzpicture}\hfill
  \begin{tikzpicture}[xscale=1,yscale=2]
    \node (b) at (2, 0) {$b'$};
    \node (c) at (4, 0) {$c'$};
  \node (d) at (6, 0) {$d'$};
  \draw (c) edge[->] node[above] {$\S$?} (b);
  \draw (c) edge[->] node[above] {$\underline\S$} (d);
    \node (bb) at (2, .25) {$\T$?};
    \node (cc) at (4, .25) {$\R$?};
  \node (dd) at (6, .25) {$\underline\T$};
  \end{tikzpicture}
  \medskip

  \medskip
  \begin{tikzpicture}[xscale=1,yscale=2]
  \node (a) at (0, 0) {$e$};
    \node (b) at (2, 0) {$f$};
    \node (c) at (4, 0) {$g$};
  \node (d) at (6, 0) {$h$};
  \draw (a) edge[->] node[above] {$\underline\S$} (b);
  \draw (c) edge[->] node[above] {$\S$?} (b);
  \draw (c) edge[->] node[above] {$\underline\S$} (d);
  \node (aa) at (0, .25) {$\underline\R$};
    \node (bb) at (2, .25) {$\T$?};
    \node (cc) at (4, .25) {$\R$?};
  \node (dd) at (6, .25) {$\underline\T$};
  \end{tikzpicture}\hfill
  \begin{tikzpicture}[xscale=1,yscale=2]
    \node (b) at (2, 0) {$f'$};
    \node (c) at (4, 0) {$g'$};
  \draw (c) edge[->] node[above] {$\S$?} (b);
    \node (bb) at (2, .25) {$\T$?};
    \node (cc) at (4, .25) {$\R$?};
  \end{tikzpicture}
  \medskip

  For the query not to be satisfied in the first (top) gadget, it must be the
  case that the fact
  $\T(b)$ is missing, as otherwise we have a query match; and the fact
  $\R(c')$ must be missing for the same reason. This gives us a probability of
  $(1-\rho)(1-\tau)$ choices. Likewise, in the (bottom) second gadget, the facts
  $\T(f)$ and $\R(g)$ must be missing, which again has a probability of
  $(1-\rho)(1-\tau)$.
  Thus, we have:

  \[\Delta^{++} = (1-\rho)(1-\tau) \Delta'\]

  Where we have $\Delta' = \deltaleft' \deltaright' - \gamma'\deltacut$, in
  which:
  \begin{itemize}
    \item $\deltaleft'$ is the total probability of possible worlds violating the query of the
      $(a,b,c,d)$-left-gadget containing the fact $\S(a, b)$ but not the facts
      $\T(b)$ and $\S(c, d)$;
\item $\deltaright'$ is the total probability of possible worlds violating the query of the
  $(a',b',c',d')$-right-gadget containing the fact $\S(c', d')$ but not the
      facts $\S(a', b')$ and $\R(c')$;
    \item $\gamma'$ is the total probability of possible wolds violating the query of the
      $(e,f,g,h)$-full-gadget containing the facts $\S(e, f)$ and $\S(g, h)$ but
      not the facts $\T(f)$ and $\R(g)$.
  \end{itemize}

  The status of the gadgets in~$\Delta'$ is:

  \medskip
  \begin{tikzpicture}[xscale=1,yscale=2]
  \node (a) at (0, 0) {$a$};
    \node (b) at (2, 0) {$b$};
    \node (c) at (4, 0) {$c$};
  \draw (a) edge[->] node[above] {$\underline\S$} (b);
  \draw (c) edge[->] node[above] {$\S$?} (b);
  \node (aa) at (0, .25) {$\underline\R$};
    \node (bb) at (2, .25) {$\notT$};
    \node (cc) at (4, .25) {$\R$?};
  \end{tikzpicture}\hfill
  \begin{tikzpicture}[xscale=1,yscale=2]
    \node (b) at (2, 0) {$b'$};
    \node (c) at (4, 0) {$c'$};
  \node (d) at (6, 0) {$d'$};
  \draw (c) edge[->] node[above] {$\S$?} (b);
  \draw (c) edge[->] node[above] {$\underline\S$} (d);
    \node (bb) at (2, .25) {$\T$?};
    \node (cc) at (4, .25) {$\notR$};
  \node (dd) at (6, .25) {$\underline\T$};
  \end{tikzpicture}
  \medskip

  \medskip
  \begin{tikzpicture}[xscale=1,yscale=2]
  \node (a) at (0, 0) {$e$};
    \node (b) at (2, 0) {$f$};
    \node (c) at (4, 0) {$g$};
  \node (d) at (6, 0) {$h$};
  \draw (a) edge[->] node[above] {$\underline\S$} (b);
  \draw (c) edge[->] node[above] {$\S$?} (b);
  \draw (c) edge[->] node[above] {$\underline\S$} (d);
  \node (aa) at (0, .25) {$\underline\R$};
    \node (bb) at (2, .25) {$\notT$};
    \node (cc) at (4, .25) {$\notR$};
  \node (dd) at (6, .25) {$\underline\T$};
  \end{tikzpicture}\hfill
  \begin{tikzpicture}[xscale=1,yscale=2]
    \node (b) at (2, 0) {$f'$};
    \node (c) at (4, 0) {$g'$};
  \draw (c) edge[->] node[above] {$\S$?} (b);
    \node (bb) at (2, .25) {$\T$?};
    \node (cc) at (4, .25) {$\R$?};
  \end{tikzpicture}
  \medskip

  It is now clear that $\Delta'$ is non-zero: the top gadgets can no longer
  contain a match of~$Q_1$, whereas the bottom gadgets have probability
  $1-\rho\sigma\tau$ of violating~$Q_1$. Thus,  $\Delta' = \rho\sigma\tau$ and
  $\Delta =  \sigma^2 (1-\rho)(1-\tau) \rho\sigma\tau$, which is
  non-zero because $0 < \rho, \tau < 1$ and $0 < \sigma \leq 1$.
  This concludes the proof of
  Lemma~\ref{lem:diseq}.
\end{proof}

Thus, we conclude from Equation~\eqref{eq:final2} that $x = 0$, which as we
explained implies $(c_1, d_1, d'_1) = (c_2, d_2, d_2')$. This establishes
Claim~\ref{clm:injective} and shows that all coefficients
$\alpha(c, d, d')$ of the Vandermonde matrix~$A$ are different, so it
is invertible.  This concludes the proof of
Theorem~\ref{thm:Qrst-hard}, and hence of 
our hardness result on weighted uniform reliability (Theorem~\ref{thm:hardgeneral_Q})
and on uniform reliability (Theorem~\ref{thm:UMC-dichotomy}).

\section{Deterministic Relations}\label{sec:deterministic}

We have shown our main result on uniform reliability
(Theorem~\ref{thm:UMC-dichotomy}). We did so by proving a more general result on
weighted uniform model counting (Theorem~\ref{thm:hardgeneral_Q}), but this
result does not cover the case of \emph{deterministic relations}, that is,
relations where the fixed probability of every fact is~$1$. In this section, we
address this issue.

\paragraph*{Case of the query~$Q_1$.}
For the query~$Q_1$, the weighted uniform model counting problem has three
parameters: the probabilities $\phi(\R)$, $\phi(\S)$, and $\phi(\T)$ of the
relations $\R$, $\S$, and $\T$. Our intractability result for~$Q_1$
(Theorem~\ref{thm:Qrst-hard}) in fact covered the case where $\phi(\S)$ may
be~$1$. As it turns out, this completely classifies the complexity of the query,
because of the following easy fact:

\begin{prop}
  \label{prp:prob1}
  Let $\phi$ be a function mapping $\R$, $\S$, and $\T$ to probabilities, and
  assume that one of $\phi(\R)$ or $\phi(\T)$ is~$1$. Then 
  $\WUR(Q_1, \phi)$ is solvable in polynomial time.
\end{prop}

\begin{proof}
  We only give the argument for when $\phi(\T) = 1$: the argument for the other
  case is analogous. Let $(I, \pi)$ be an input instance. First remove all
  facts $\S(a, b)$ from~$I$ where $I$ does not contain the fact $\T(b)$: this
  clearly does not change the answer to the problem as such facts can never be
  part of a match to the query. Let $(I', \pi')$ be the result of this process,
  with $\pi'$ being the restriction of~$\pi$ to~$I'$. Now, observe that every match of the query
  $Q_1': \R(x), \S(x, y)$ in~$I$ translates to a match of the query~$Q_1$ on the same
  choice of~$x, y$; and conversely it is obvious that any match of~$Q_1$
  translates to a match of~$Q_1'$. Hence, every possible world of~$(I, \pi)$ has
  a match of~$Q_1$ iff it has a match of~$Q_1'$. This implies that we can solve
  $\PQE_{r,s,1}(Q_1)$ on~$I'$, hence on~$I$, by solving $\PQE(Q_1')$, and this
  is tractable by the result of Dalvi and Suciu~\cite{dalvi2007efficient} (Theorem~\ref{thm:DS-dichotomy})
  because $Q_1'$ is a hierarchical self-join-free CQ.
\end{proof}

Thus, we have the following classification for weighted uniform model counting
for the query~$Q_1$, which was conjectured in the conference version of this
paper \cite[Conjecture~7.4] {DBLP:conf/icdt/AmarilliK21}:

\begin{prop}\label{prop:rst}
  Let $\phi$ be a function mapping $\R$, $\S$, and $\T$ to probabilities. If
  $\phi(\R) < 1$ and $\phi(\T) < 1$, then $\WUR(Q_1, \phi)$ is \#P-hard;
  otherwise, it is solvable in polynomial time.
\end{prop}

\paragraph*{Case of general queries.}
For arbitrary CQs without self-joins, deterministic relations
certainly have an impact on the complexity of weighted uniform
reliability, and we cannot hope that Theorem~\ref{thm:hardgeneral_Q}
generalizes as-is: for instance, if all relations have
probability~$1$, then the weighted uniform reliability problem is
equivalent to evaluating a fixed CQ on non-probabilistic data, and hence,
in polynomial time. More generally, our reduction
from~$Q_1$ (Lemma~\ref{lemma:reduction_detailed}) does not work
immediately in the same way, as a simple example illustrates:

\begin{exa}
  Consider the query $Q: \R(x), \S_1(x, y), \S_2(y, z), \T(z)$, and the function
  $\phi$ mapping $\R$ and $\T$ to $\frac12$ and $\S_1$ and $\S_2$ to~$1$. It is
  clear that $\WUR(Q, \phi)$ is \#P-hard by a simple reduction from weighted
  uniform reliability for $Q_1$, replacing every fact $\S(a, b)$ by two facts
  $\S_1(a, d), \S_2(d, b)$ for some fresh~$d$. Yet, we cannot find a match
  of~$Q_1$ in~$Q$ by renaming variables as in the proof of
  Lemma~\ref{lemma:reduction_detailed}.
\end{exa}

Fortunately, there is a classification of the hardness of
probabilistic query evaluation for Boolean CQs without self-joins in
\cite[Theorem 8]{dalvi2007efficient} in the case where some relations
are deterministic, giving a pattern of relations that characterizes
intractability. We can re-use this classification and show that the
same dichotomy applies for weighted uniform reliability: an
intractable query for PQE is also hard under the assumption that all
tuples in all probabilistic relations have a common, fixed
probability.
Here is the criterion used in their dichotomy:

\begin{defi}[From \cite{dalvi2007efficient}, Theorem~8]
  Let $Q$ be a CQ without self-joins, and $\phi$ be a function mapping the
  relations of~$Q$ to probabilities in $(0, 1]$. A \emph{hardness pattern}
  of~$Q$ and~$\phi$ is a
  sequence of relations $\R_0, \ldots, \R_{k+1}$, with corresponding atoms
  $U_i(\tup{a}_i)$ in~$Q$, such that:
  \begin{itemize}
    \item $\phi(\R_0) < 1$ and $\phi(\R_{k+1}) < 1$.
    \item There is a variable $x$ occurring in $\tup{a}_0$ and $\tup{a}_1$ but
      not in $\tup{a}_{k+1}$.
    \item There is a variable $y$ occurring in $\tup{a}_k$ and $\tup{a}_{k+1}$ but
      not in $\tup{a}_0$.
    \item For all $i=1,\dots, k-1$ we have that
      $\tup{a}_i \cap \tup{a}_{i+1} \not\subseteq \tup{a}_0 \cup
      \tup{a}_{k+1}$ (that is, every two internal consecutive atoms have a
      common variable that is in neither $\tup{a}_0$ nor
      $\tup{a}_{k+1} $), where we abuse notation and see tuples as
      sets.
  \end{itemize}
\end{defi}

We show the following dichotomy result. The result completely classifies the
complexity of the weighted uniform reliability problem for CQs without
self-joins when deterministic relations are allowed, at the expense of a
somewhat more complex classification:
\begin{thm}
  \label{thm:deterministic}
  Let $Q$ be a CQ without self-joins, and $\phi$ be a function mapping
  the relations of~$Q$ to probabilities in $(0, 1]$. If $Q$ and $\phi$
  do not have a hardness pattern, then $\WUR(Q, \phi)$ is solvable in
  polynomial time; otherwise, it is \#P-hard.
\end{thm}

\begin{proof}
  If $Q$ and $\phi$ do not have a hardness pattern, then we know by
  \cite[Theorem 8]{dalvi2007efficient} that PQE for~$Q$ is PTIME under
  the restriction that tuples of relations mapped to~$1$ by~$\phi$
  have probability~$1$. Hence, as weighted uniform reliability for~$Q$
  under~$\phi$ is a special case of PQE for~$Q$ under this
  requirement, it is solvable in polynomial time.

  Now, if $Q$ and $\phi$ have a hardness pattern, we show that $\WUR(Q, \phi)$
  is \#P-hard. We first note that, if $k=1$, then the definition of a hardness
  pattern implies that we can conclude as in
  Lemma~\ref{lemma:reduction_detailed}. Specifically, in this case, we have
  relations $\R_0, \R_1, \R_2$, with $\phi(\R_0) < 1$, $\phi(\R_2) < 1$, and a
  variable $x$ occurring in $\R_0$ and $\R_1$ but not $\R_2$, and a variable $y$
  occurring in $\R_1$ and $\R_2$ but not $\R_0$, and we conclude like in
  Lemma~\ref{lemma:reduction_detailed}. Hence, to simplify the presentation, we
  assume that $k>1$ from now on.

  We show hardness by reducing from $\WUR(Q_1, \phi_1)$ for
  some~$\phi_1$ that we will define later, analogously to the proof of
  Lemma~\ref{lemma:reduction_detailed}, and with some inspiration from
  the proof of Dalvi and Suciu~\cite[Theorem
  8]{dalvi2007efficient}. To define $\phi_1$, we partition the
  variables of~$Q$ into four sets, again abusing set notation to apply
  to tuples:
  \begin{itemize}
    \item the set
  $S_x \colonequals \tup{a}_0 \setminus \tup{a}_{k+1}$ (so
  including~$x$);
\item the set
  $S_y \colonequals \tup{a}_{k+1} \setminus \tup{a}_0$ (so
  including~$y$);
\item the set
  $S_C \colonequals \bigcup_{1 \leq i \leq k-1} \tup{a}_i \cap
  \tup{a}_{i+1} \setminus (\tup{a}_0 \cup \tup{a}_{k+1})$ (by
  the fourth condition in the definition of a hardness pattern, and as $k>1$,
      this set is non-empty);
\item and the set
  $S_0$ of all other variables (including in particular
  $\tup{a}_0 \cap \tup{a}_{k+1}$, which may be empty).
  \end{itemize}
  Note that the sets
  $S_x$, $S_y$, and $S_C$ is non-empty, but $S_0$
  may be empty.

  Following this partition of variables, we partition the relations of~$Q$ into
  five kinds depending on the variables that their corresponding atom contain
  (we ignore the constants that the atoms may also contain):
  \begin{itemize}
    \item $0$-relations, which contain only variables of~$S_0$, including any
      relations that contain no variables at all; it may be the case that there
      are no $0$-relations.
    \item $x$-relations, containing only variables of~$S_x \cup S_0$ and at
      least one
      variable of~$S_x$; there is at least one $x$-relation, namely~$\R_0$.
    \item $y$-relations, containing only variables of~$S_y \cup S_0$ and at
      least one
      variable of~$S_y$; there is at least one $y$-relation, namely~$\R_{k+1}$.
    \item $xy$-relations, containing only variables of~$S_x \cup S_y \cup S_0$
      and at least one variable of~$S_x$ and at least one variable of~$S_y$; it
      may be the case that there are no $xy$-relations.
    \item $C$-relations, containing a variable of~$S_C$; there is at least $k$
      $C$-relations, namely, the relations $\R_1, \ldots, \R_{k-1}$, so as $k>1$
      there is at least one such relation.
    \end{itemize}
    As an example, in the following CQ, assume that $\phi(\mathsf{R}) < 1$ and $\phi(\mathsf{U}) <
1$.
    \[\mathsf{P}(x, y), \mathsf{R}(x, y, z),
  \mathsf{S}(x, w, w'), \mathsf{T}(w, w', z, x, u), \mathsf{U}(u, z),
  \allowbreak \mathsf{V}(x, z,
  u), \mathsf{V}'(x, z, w'), \mathsf{W}(t), \mathsf{W}'(t, z)\]
\begin{itemize}
  \item The sequence $\mathsf{R}, \mathsf{S}, \mathsf{T}, \mathsf{U}$ is a
  hardness pattern: the variable $x$ occurs in the $\mathsf{R}$- and
  $\mathsf{S}$-atoms but not in the~$\mathsf{U}$-atom, the variable $u$ occurs
  in the $\mathsf{T}$- and $\mathsf{U}$-atoms but not in the $\mathsf{R}$-atom,
  the variable $w$ occurs in the $\mathsf{S}$- and $\mathsf{T}$-atoms but not in
  the $\mathsf{R}$- or $\mathsf{U}$-atoms.
\item The set $S_x$ contains $\{x, y\}$, the set $S_y$ contains
  $\{u\}$, the set $S_C$ contains $\{w, w'\}$, and the set $S_0$
  contains all other variables.
\item The relations $\mathsf{P}$ and $\mathsf{R}$ are $x$-relations,
  the relation $\mathsf{U}$ is a $y$-relation, the relation
  $\mathsf{V}$ is an $xy$-relation, the relations $\mathsf{S}$ and
  $\mathsf{T}$ and $\mathsf{V}'$ are $C$-relations, and the relations
  $\mathsf{W}$ and $\mathsf{W}'$ are $0$-relations.
\end{itemize}
  
    Next, we show a reduction from $\WUR(Q_1, \phi_1)$ for some hard
    case of $\phi_1$.  We define $\phi_1$ by mapping the relation $\R$
    to the product of the $\phi$-values of all $x$-relations (which is
    strictly smaller than $1$, because $\phi(\R_0) < 1$), mapping the relation~$\T$ to the
    product of the $\phi$-values of all $y$-relations (which is strictly smaller than $1$,
    because $\phi(\R_{k+1}) < 1$), and mapping the relation~$\S$ to
    the product of the $\phi$-values of all~$xy$-relations and
    $C$-relations (which may be $1$ or smaller than $1$).

  Let $I_1$ be an input instance for $\WUR(Q_1, \phi_1)$. As in the proof of
  Lemma~\ref{lemma:reduction_detailed}, we call a \emph{match} of~$Q_1$ in~$I_1$
  a pair $(a,b)$ of constants such that $I_1$ contains all of $\R(a)$,
  $\S(a,b)$ and $\T(b)$, and we assume without
  loss of generality that every fact of~$I_1$ is part of a match.

  We construct from~$I_1$ the instance~$I$ for~$Q$ as follows: for every match
  $(a, b)$ in~$I_1$, we add to~$I$ a copy of the query~$Q$ where all variables of~$S_x$
  are replaced by~$a$, all variables of~$S_y$ are replaced by~$b$, all variables
  of~$S_0$ are replaced by a fixed constant~$c$ (always the same), and all
  variables of~$S_C$ are replaced by a fresh constant~$c_{a,b}$. Note that this
  may include the same fact multiple times, in which case it is only added once. To
  be more precise, we can equivalently state the construction as
  inserting the following facts into $I$, where we describe the facts in term of
  the elements that replace variables (any constants used in the CQ are left
  as-is in the atoms):
  \begin{itemize}
    \item For each $0$-relation, one fact of the $0$-relation where all
      variables (if any) are replaced by the fixed constant~$c$;
    \item For each element~$a$ of~$I_1$ involved in a match $(a, b)$ for
      some~$b$, and for each $x$-relation, one fact of that
      relation where variables of~$S_0$ (if any) are replaced by~$c$ and
      variables of~$S_x$ (including~$x$) are replaced by~$a$;
    \item For each element~$b$ of~$I_1$ involved in a match~$(a, b)$ for
      some~$a$, and for each $y$-relation, one fact of that
      relation where variables of~$S_0$ (if any) are replaced by~$c$ and
      variables of~$S_y$ (including~$y$) are replaced by~$b$;
    \item For each match $(a, b)$ of~$I_1$, and for each $xy$-relation, one fact of that
      relation where variables of~$S_x$ are replaced by~$a$, variables of~$S_y$
      are replaced by~$b$, and variables of~$S_0$ are replaced by~$c$;
    \item For each match $(a, b)$ of~$I_1$, and for each $C$-relation, one fact of that
      relation where variables of~$S_C$ are replaced by $c_{a,b}$, variables
      of~$S_x$ (if any) are replaced by~$a$, variables of~$S_y$ (if any) are
      replaced by~$b$, and variables of~$S_0$ (if any) are replaced by~$c$.
  \end{itemize}
  This construction is clearly in polynomial time.

  Continuing from our example, we would produce the facts $\mathsf{W}(c)$,
  $\mathsf{W}'(c, c)$, the facts $\mathsf{P}(a, a)$ and $\mathsf{R}(a, a, c)$
  for each $a$ in a match $(a, b)$, the fact $\mathsf{U}(b, c)$ for each $b$ in
  a match $(a, b)$, and the facts $\mathsf{S}(a, c_{a,b}, c_{a,b})$,
  $\mathsf{T}(c_{a,b}, c_{a,b}, c, a, b)$, $\mathsf{V}'(a, c, c_{a, b})$, and
  $\mathsf{V}(a, c, b)$ for each match $(a, b)$.
  Hence, we add the following for every match $(a,b)$:
  \[\mathsf{P}(a, a)\, \mathsf{R}(a, a, c)\,
    \mathsf{S}(a, c_{a,b}, c_{a,b})\, \mathsf{T}(c_{a,b}, c_{a,b}, c, a, b)\, \mathsf{U}(b, c)\,
  \mathsf{V}(a, c, b)\, \mathsf{V}'(a, c, c_{a, b})\, \mathsf{W}(c)\, \mathsf{W}'(c, c)\]
  
  Now, as in the proof of Lemma~\ref{lemma:reduction_detailed}, we study the
  possible worlds of~$I$ that satisfy~$Q$. We first note that, for $0$-relations,
  we created only one fact for this relation (containing only the constant~$c$
  and possibly constants used in the CQ),
  and this fact must be kept to have a query match, so the probability of
  satisfying~$Q$ in~$I$ includes a factor $\prod
  \phi(\R)$ across all the $0$-relations to keep their facts, and we can
  therefore focus on the possible worlds where all these facts
  are kept. We then define a possible world $J$ of~$I$ by choosing a
  possible world $J_1$ of~$I_1$ and:
  \begin{itemize}
    \item For each fact $\R(a)$ of~$I_1$:
      \begin{itemize}
        \item If $\R(a)$ is in~$J_1$, we keep the facts of all $x$-relations
          involving~$a$;
        \item Otherwise, we do not keep all these facts, i.e., we discard at
          least one of these facts.
      \end{itemize}
    \item For each fact $\S(a, b)$ of~$I_1$:
      \begin{itemize}
        \item If $\S(a, b)$ is in~$J_1$, we keep the facts of all $xy$-relations
          involving~$a$ and~$b$, and all facts of~$C$-relations involving
          $c_{ab}$;
        \item Otherwise, we do not keep all these facts, i.e., we discard at
          least one of these facts.
      \end{itemize}
    \item For each fact $\T(b)$ of~$I_1$:
      \begin{itemize}
        \item If $\T(b)$ is in~$J_1$, we keep the facts of all $y$-relations
          involving~$b$;
        \item Otherwise, we do not keep all these facts, i.e., we discard at
          least one of these facts.
      \end{itemize}
    \item We keep all facts of~the $0$-relations.
    \end{itemize}
  Now, if our choice of~$J_1 \subseteq I_1$ satisfies~$Q_1$, then the
  facts retained in~$J$ witness that $J$ satisfies~$Q$. Conversely, if
  we have a possible world~$J$ of~$I$ that satisfies~$Q$, we must argue
  that the corresponding possible world~$J_1$ of~$I_1$
  satisfies~$Q_1$. Intuitively, the challenge is to argue that the facts used to
  satisfy~$Q$ in~$J$ can be chosen from one single match of~$Q_1$ in~$I_1$ such
  that all facts created from this match in the construction were kept in~$J$.

  To show this, we first observe that, in the mapping from~$Q$ to~$J$ witnessing
  that $Q$ is satisfied in~$J$, the variables of~$S_0$ must be mapped to the
  constant~$c$, because $c$ is the only element used to create facts at the
  positions corresponding to variables of~$S_0$. Now, consider the matches in~$J$ of the atoms
  for relations $\R_0, \ldots, \R_{k+1}$ of~$Q$. The matches of~$\R_0$ and
  $\R_{k+1}$ give us a value $a$ for all variables in~$S_x$: by construction of
  the $\R_0$--facts in~$I$, all variables of the set~$S_x$ must be mapped to the
  same element, which is some element $a$ for a match $(a, b)$ in~$I_1$. We
  obtain a value $b$ for all variables of~$S_y$ in the same way. This
  ensures that all the facts for~$x$-relations for this value of~$a$ must
  be kept in~$J$, and likewise for the facts for~$y$-relations for this
  value of~$b$, and also for the facts for~$xy$-relations for this value
  of~$a$ and~$b$.

  It remains to study to which elements in~$J$ the variables at positions
  in~$S_C$ can be mapped by our match of~$Q$, and to show that the facts
  for~$C$-relations for the match $(a, b)$ of~$Q_1$ in~$I_1$ must be kept
  in~$J$. We do so by showing that all variables of~$S_C$ were in fact mapped to
  the element~$c_{a,b}$ determined from the~$a$ and~$b$ already defined. To do
  this, as we know that $x$ is in $\R_1$, we know that the
  $\R_1$-atom of~$Q$ is matched to a fact involving~$a$ and some $c_{a',b'}$
  for the variables of
  $\tup{a}_1 \cap \tup{a}_2 \setminus (\tup{a}_0 \cup \tup{a}_{k+1})$ which is a
  non-empty subset of~$S_C$:
  by construction, it must be $c_{a, b'}$ for some~$b'$. But then, the
  $\R_2$-atom is matched to a fact involving this~$c_{a, b'}$ and some
  $c_{a', b''}$ for the variables of
  $\tup{a}_2 \cap \tup{a}_3 \setminus (\tup{a}_0 \cup \tup{a}_{k+1})$:
  by construction we must have $a = a'$ and $b' = b''$. Repeating the
  argument across the path, we show that all variables of $S_C$ are
  mapped to the same $c_{a, b'}$, and the only way to avoid a
  contradiction in the end (given that $y$ is in $\R_k$ and
  $\R_{k+1}$) is that $b' = b$. Therefore, it is indeed the case that the match
  of~$Q$ in~$J$ maps all variables of~$S_C$ to~$c_{a,b}$, and it witnesses that all the facts
  for $C$-relations for~$(a, b)$ must be kept.

  To summarize, the match of~$Q$ in~$J$ must map all
  variables of~$S_x$ to~$a$, all variables of~$S_y$ to~$b$, and all
  variables of~$S_C$ to~$c_{a,b}$, for some choice of~$(a, b)$ which is a match
  of~$Q_1$ in~$I_1$. Further, we
  know that all $x$-facts involving~$a$, all $y$-facts involving~$b$,
  all $xy$-facts involving $a$ and~$b$, all $C$-facts
  involving~$c_{a,b}$, and all $0$-facts, were kept in~$J$. Thus, the
  choice of $J_1 \subseteq I_1$ that corresponds to our choice of~$J \subseteq
  I$ must satisfy~$Q_1$.

  Concluding the proof as in Lemma~\ref{lemma:reduction_detailed}, we
  establish that the probability that $Q$ is satisfied in~$I$ is
  exactly $\prod \phi(\R)$ for all $0$-relations $\R$, times the
  probability that $Q_1$ is satisfied in~$I_1$. This concludes the
  reduction. Thus, our hardness result follows from our hardness
  result on weighted uniform reliability for~$Q_1$
  (Theorem~\ref{thm:hardgeneral_Q}).
\end{proof}

\section{Concluding Remarks}
\label{sec:conc}
While query evaluation over TIDs has been studied for over a decade,
the basic case of a uniform distribution, namely uniform reliability,
had been left open. We have settled this open question for the class
of CQs without self-joins, and shown a dichotomy on computational
complexity of counting satisfying database subsets: this task is
tractable for hierarchical queries, and \#P-hard otherwise.  Our precise
result is more general and applies to weighted uniform reliability,
where each relation is associated with a probability that is attached
to all of its facts.

One immediate question
for future research is whether our results could extend to
more general query classes. The obvious challenge is to extend to CQs with
self-joins and UCQs with self-joins, and try to match the known dichotomy for
non-uniform probabilities~\cite{dalvi2012dichotomy}. Following our
work, considerable progress in this direction has been done recently
by Kenig and Suciu~\cite{DBLP:journals/corr/abs-2008-00896}, which
addresses the case of PQE with probabilities of $\frac 12$ and~$1$, and
leaves open the case of uniform reliability for arbitrary UCQs. The same study
could be undertaken for the general class of queries closed under homomorphisms,
following the recent dichotomy on PQE for such queries (on binary signatures)
in~\cite{amarilli2022dichotomy}.

\section*{Acknowledgment}
The work of Antoine Amarilli was partially supported by the ANR project EQUUS
ANR-19-CE48-0019 and by the Deutsche Forschungsgemeinschaft (DFG, German Research Foundation) – 431183758.
The work of Benny Kimelfeld was supported by the Israel Science
  Foundation (ISF), Grant 768/19, and the German Research Foundation (DFG) Project 412400621 (DIP program).

\bibliography{main}
\bibliographystyle{alphaurl}

\end{document}